\pdfoutput=1
\documentclass[a4paper,fleqn,usenatbib]{mnras}
\usepackage{newtxtext,newtxmath}
\usepackage[T1]{fontenc}
\usepackage{ae,aecompl}
\usepackage{longtable}
\usepackage{threeparttable}
\usepackage{natbib}
\usepackage{adjustbox}
\usepackage{graphicx}
\bibpunct{(}{)}{;}{a}{}{,}
\usepackage{booktabs}
\usepackage[T1]{fontenc}
\usepackage{aecompl}
\usepackage{multicol}
\usepackage{multirow}
\usepackage{placeins}
\usepackage{color}

\title[]{An investigation of C, N and Na abundances in red giant stars of the Sculptor dwarf spheroidal galaxy}
\author[Salgado et al.]{
C. Salgado,$^{1}$\thanks{Contact e-mail: \href{mailto:carolina.salgado@anu.edu.au}{carolina.salgado@anu.edu.au}}
G. S. Da Costa,$^{1}$
J. E. Norris,$^{1}$
D. Yong$^{1}$
\\
$^{1}$Research School of Astronomy and Astrophysics, Australian National University, Canberra, ACT 2611, Australia\\
}

\date{Accepted XXX. Received YYY; in original form ZZZ}

\pubyear{2018}

\begin{document}
\label{firstpage}
\pagerange{\pageref{firstpage}--\pageref{lastpage}}
\maketitle

\begin{abstract}
The origin of the star-to-star abundance variations found for the light elements in Galactic globular clusters (GGCs) is not well understood, which is a significant problem for stellar astrophysics. While the light element abundance variations are very common in globular clusters, they are comparatively rare in the Galactic halo field population. However, little is known regarding the occurrence of the abundance anomalies in other environments such as that of dwarf spheroidal (dSph) galaxies.
Consequently, we have investigated the anti-correlation and bimodality of CH and CN band strengths,
which are markers of the abundance variations in GGCs, in the spectra of red giants in the Sculptor dwarf spheroidal galaxy.  Using spectra at the Na~D lines, informed by similar spectra for five GGCs (NGC 288, 1851, 6752, 6809 and 7099), we have also searched for any correlation between CN and Na in the Sculptor red giant sample.  Our results indicate that variations analogous to those seen in GGCs are not present in our Sculptor sample. Instead, we find a weak positive
correlation between CH and CN, and no correlation between Na and CN.  We also reveal a deficiency in [Na/Fe] for the Sculptor stars relative to the values in GGCs, a result which is consistent with previous work for dSph galaxies. The outcomes reinforce the apparent need for a high stellar density environment to produce the light element abundance variations.

\end{abstract}


\section{Introduction}
\label{s_intro}

The study of Galactic globular clusters (GGCs) allows the investigation of many aspects of stellar evolution, and of the dynamics and chemical conditions at the moment when the host galaxy formed. 
However, detailed photometric and spectroscopic studies have demonstrated that GGCs possess a much more complex star formation history than previously thought. One of the main unsolved problems is the origin of the star-to-star abundance variation in the light elements (C, N, O, Na, Mg and Al) and in helium that is seen in most, if not all, globular clusters. While approximately half of the stars in a typical cluster have abundance patterns similar to halo field stars, the other half of the stars in the cluster are depleted in carbon, oxygen and magnesium, and enhanced in nitrogen, sodium, aluminium \citep[e.g.,][]{OsbornW1971,norris1981abundance,kraft1994,gratton2001,gratton2004abundance,salaris2006primordial}. 
While the evidence for the light element abundance variations is provided, inter alia, directly by spectroscopic determinations, the evidence for helium abundance variations comes primarily from the existence of split or multiple main sequences observed in high precision {\it HST}-based color-magnitude diagrams \citep[e.g. NGC6752,][]{milone2013wfc3}, and from interpretations of horizontal branch morphologies \citep[e.g.][]{dantona2005}, though direct spectroscopic evidence does exist \citep[e.g.][]{marino2014}.  See also \citet{milone2018}.

The effects of H-burning at high temperatures provide a clue to the origin of the observed abundance variations, as the abundances of light elements can be altered by the simultaneous action of p-capture reactions in the CNO, NeNa and MgAl chains \citep{denisenkov1989,langer1993, prantzos2007light}. The temperatures required for these processes to occur are $\geq {20 \times 10}^6$ K  for the NeNa cycle and $\geq {70 \times 10}^6$ K for the MgAl cycle. However, such temperatures are not reached in the interior of current GGC stars requiring the anomalies to come from an earlier generation or generations of stars. In order to explain these light element abundance variations, several types of stars have been proposed as possible polluters. The most popular are intermediate-mass asymptotic giant branch (AGB) stars~\citep{cottrell1981,dantona1983,ventura2001} and/or super-AGB stars \citep{pumo2008super,ventura2011,d2016single}, supermassive stars \citep{denissenkov2013supermassive}  and fast rotating massive stars (FRMS) \citep{norris2014, maeder2006, decressin2007fast, decressin2009cno}. These stars allow a hot H-burning environment, a mechanism to bring the processed material to the surface (convection and rotational mixing for AGBs and FRMS, respectively), and a way to release this material into the intra-cluster medium at a velocity sufficiently low to avoid escaping from the cluster potential well. \citet{renzini2015hubble} test, discuss and summarize several scenarios and reveal the successes and deficiencies of each of them based on the evidence provided by the results of the Hubble Space Telescope (HST) UV Legacy Survey of GGCs  \citep{piotto2015hubble}.  See also \citet{bastian2018multiple}. 

The  abundance anomalies have been found in all evolutionary phases of GGC stars from the main sequence to the giant branch. Specifically, the presence of abundance anomalies on the main sequence (MS) of several GGCs, for example NGC~104, NGC~288, NGC~6205, and NGC~6752 \citep[e.g.,][]{smith1982cyanogen, sneden1994oxygen, lee2004infrared, gratton2004abundance, carretta2010properties} requires that the anomalies come from a previous generation or generations.  We note, however, that star-to-star variations in the heavy elements (e.g. Ca and Fe) are not common in GGCs, and are  instead restricted to a small number of clusters such as $\omega$ Centauri, M22 and M54.  

The study of chemical abundance patterns is not restricted to GGCs: for example, their study in individual metal-poor stars provides clues about element formation and evolution in the universe. In particular, carbon and nitrogen play a crucial role in understanding the evolution of galaxies because they are produced by different mechanisms and by stars of a wide range of mass. Many studies show that the strength of the CN molecular bands in the spectra of red giant stars is a good indicator of [N/Fe], while the CH-band provides a measure of [C/Fe] \citep[e.g.,][]{smith1996cno}. Carbon and nitrogen have been studied in different environments.  For example, in GGC red giants, spectroscopic studies of CN and CH band strengths have shown that there is generally an anti-correlation between CN- and CH-band strength, coupled with a bimodality of CN-band strength that allows the classification of stars as CN-strong or CN-weak, \citep[e.g.,][]{norris1981abundance, norris1984anticorrelation, pancino2010}.  Spectrum synthesis calculations have demonstrated that these CN- and CH-band strength variations are a direct consequence of differences in C and N abundances, with C depleted and N enhanced in the so-called `second generation' (CN-strong) stars as compared to the Galactic halo-like abundances in the `first generation'  (CN-weak) stars. 

As regards dSph galaxies, \citet{shetrone2013carbon} analysed CN and CH molecular band strengths in the spectra of 35 red giants in the Draco dSph. They found little evidence for any spread in CN-band strength at a given luminosity, and, in contrast to the anti-correlation seen in GGCs, the CN and CH band strengths were found to be generally correlated.  Using spectrum syntheses they showed that the carbon abundances decrease with increasing luminosity consistent with the expectations of evolutionary mixing, a phenomenon that is also seen in field stars and globular clusters in the halo.  \citet{shetrone2013carbon} also found evidence for an intrinsic spread in [C/Fe] at fixed [Fe/H] and/or M$_{bol}$ in their sample, which they intrepreted as having at least a partial primordial origin.
 
 \citet{kirby2015carbon} also studied the carbon abundances of a large sample of red giant stars in both GGCs and dSphs seeking to understand the relation between the dSphs, the Galactic halo and their chemical enrichment. For non-carbon enhanced stars, i.e., those with [C/Fe] $\leq$ 0.7, \citet{kirby2015carbon} compared the trend of [C/Fe] versus [Fe/H] for dSphs with the trend for Galactic halo stars, finding that the `knee' in [C/Fe] occurs at a lower metallicity in the dSphs than in the Galactic halo. The knee corresponds to the metallicity at which SNe Ia begin to contribute to the chemical evolution, and the difference in the location of the knee caused \citet{kirby2015carbon} to suggest that SNe Ia activity is more important at lower metallicities in dSphs than it is in the halo.
 
In a similar fashion \citet{lardo2016carbon} reported the results of a study of carbon and nitrogen abundance ratios, obtained from CH and CN index measurements, for  94 red giant branch (RGB) stars in the Sculptor (Scl) dSph. The results indicate that [C/Fe] decreases with increasing luminosity across the full metallicity range on the Scl red giant branch.  More specifically, the measurements of [C/Fe] and [N/Fe] are in excellent agreement with theoretical model predictions \citep{stancliffe2009depletion} that show  the red giants experience a significant depletion of carbon after the first dredge-up.   \citet{lardo2016carbon} also reported the discovery of two carbon-enhanced metal-poor (CEMP) stars in Scl, both of which show an excess of barium consistent with $s$-process nucleosynthesis. 

Studies of carbon abundances have also been made in ultra-faint dwarf galaxies: for example, \citet{norris2010chemical} present carbon abundances for red giant stars in the Bo\"otes~I and Segue~1 systems. They show that the stars in these ultra-faint dwarf galaxies that have [Fe/H] $\leq$ --3.0 exhibit a relation between [C/Fe] and [Fe/H] similar to that seen for Galactic halo stars.  Further,  \citet{norris2010chemical} also found a carbon-rich and extremely metal-poor star ([C/Fe] = +2.3, [Fe/H]= --3.5) in Segue 1 that is similar to the CEMP stars found in the Galactic halo.

However, it is not possible to directly relate any of these findings to the GGCs abundance anomalies problem without information on the Na abundance in the stars: the GGC characteristic signature of Na-enhancement coupled with 
C-depletion and N-enhancement needs to be investigated to decide if the C, N-variations in Scl and other dSphs are merely the result of evolutionary mixing on the RGB, or arise from other nucleosynthetic processes unrelated to that for the GGC abundance anomalies.  
A further signature of the abundance variations in GGCs is the anti-correlation between sodium and oxygen abundances, which is seen at all evolutionary phases including the main sequence \citep{gratton2001and}. The large spread in Na and O in GGCs \citep[e.g.,][]{carretta2009anticorrelation, carretta2016spectroscopic} further emphasizes that GGCs are not simple stellar populations.  This signature also results in a correlation between sodium abundances and CN-band strengths, in which CN-strong red giant stars (enhanced in N, depleted in C) are also enhanced in Na \citep{cottrell1981,  norris1985sodium}.

In this context we note that \citet{shetrone2003vlt} has studied [Na/Fe] values for small samples of red giants in the Carina, Sculptor, Fornax and Leo~I  dwarf spheroidal galaxies.  The results show that the dSph stars have low [Na/Fe] values compared to halo stars with similar [Fe/H].  Further, \citet{geisler2005sculptor} measured [Na/Fe] for small number of red giants in Sculptor and found a similar result: the dSph stars were deficient in [Na/Fe] compared to `first generation' GGC and Galactic halo field stars.  \citet{letarte2010} 
and \citet{lemasle2014} have shown that this is also the case in the Fornax dSph.  However, in contrast, \citep{aoki2009chemical} found [Na/Fe]$\sim$ 0.0 for five stars in Sextans dSph galaxy, although they also found one star with a very low abundance of sodium: [Na/Fe]$\sim$ --0.7.  \citet{norris2017populations} studied the chemical enrichment of the Carina dSph via an abundance study of 63 RGB stars.  They showed that the Carina stars also have lower [Na/Fe] compared to GGC red giants, and most significantly in the current context, they found no strong evidence in the Carina stars for the Na-O anti-correlation present in GGCs \citep{norris2017populations}.

As regards the presence of GGC abundance anomalies in the halo field, \citet{martell2011building} present results from spectra of 561 halo red giants. They searched for the anti-correlated CN- and 
CH-band strength behaviour seen in GGCs red giants, but found that just  $\sim$3\% of the sample exhibited band strengths similar to those of second generation GGC stars. These halo field stars could be second generation globular cluster stars that have either escaped from their parent cluster or which originated in now disrupted GGCs.  Using theoretical models of globular cluster formation with two generations of stars, \citet{martell2011building} interpreted their results as suggesting that escape from, and dissolution of, globular clusters could contribute significantly to the stellar population of the Galactic halo; their estimate is approximately $\sim$17\% of the currently mass of the stellar halo originated in this way.  
Similarly, \citet{schiavon2016chemical} report the discovery in the APOGEE survey \citep{majewski2016}
of a population field stars in the inner Galaxy with high [N/Fe] that is correlated with [Al/Fe] and anti-correlated with [C/Fe] in similar fashion to that for the second generation population in GGCs.  If these stars have an origin as former members of disrupted GGCs then the total mass of such clusters must substantially exceed that of the surviving GGC system \citep{schiavon2016chemical}, although the numbers may also indicate that the stars have a separate origin to that of the GGC stars \citep{schiavon2016chemical}.

The aim of this work is to investigate if the abundance variations of the light elements seen in globular clusters are also present among the stars of the Sculptor dwarf spheroidal galaxy, in order to help constrain the origin of the anomalies. Sculptor is a well-studied dSph satellite galaxy of the Milky Way.  It has  $M_V\approx -11.2$, is strongly dominated by dark matter \citep{walker2007velocity}, and lies at a distance of $86\pm 5$ kpc \citep{pietrzynski2008araucaria} from the Sun.  The stellar population of Scl is dominated by old (age $>10$ Gyr) metal-poor stars, and no significant star formation has occurred in the system for at least $\approx 6$ Gyr \citep{de2012star}. In this paper we investigate the strengths of the CN- and CH-bands in the spectra of Sculptor red giants in order to test if the anti-correlation and bimodality seen in GGCs are also present in Sculptor.  As an additional constraint we have also estimated sodium abundances and explored the extent of any Na/CN correlation similar to that seen in the GGCs.

Our paper is arranged as follows. In section \ref{s_obs} we present the observations, data reduction and definition of the indices and their measurement; in section \ref{metallicities_procedure} we outline our analysis procedure to derive overall metallicities; in section \ref{results cn ch na} we present and discuss the results, and in section \ref{s_concl} we summarize our findings and present our conclusions.

\section{Observations and data reduction}
\label{s_obs}

\begin{figure}
\begin{center}
\includegraphics[width=8.8cm]{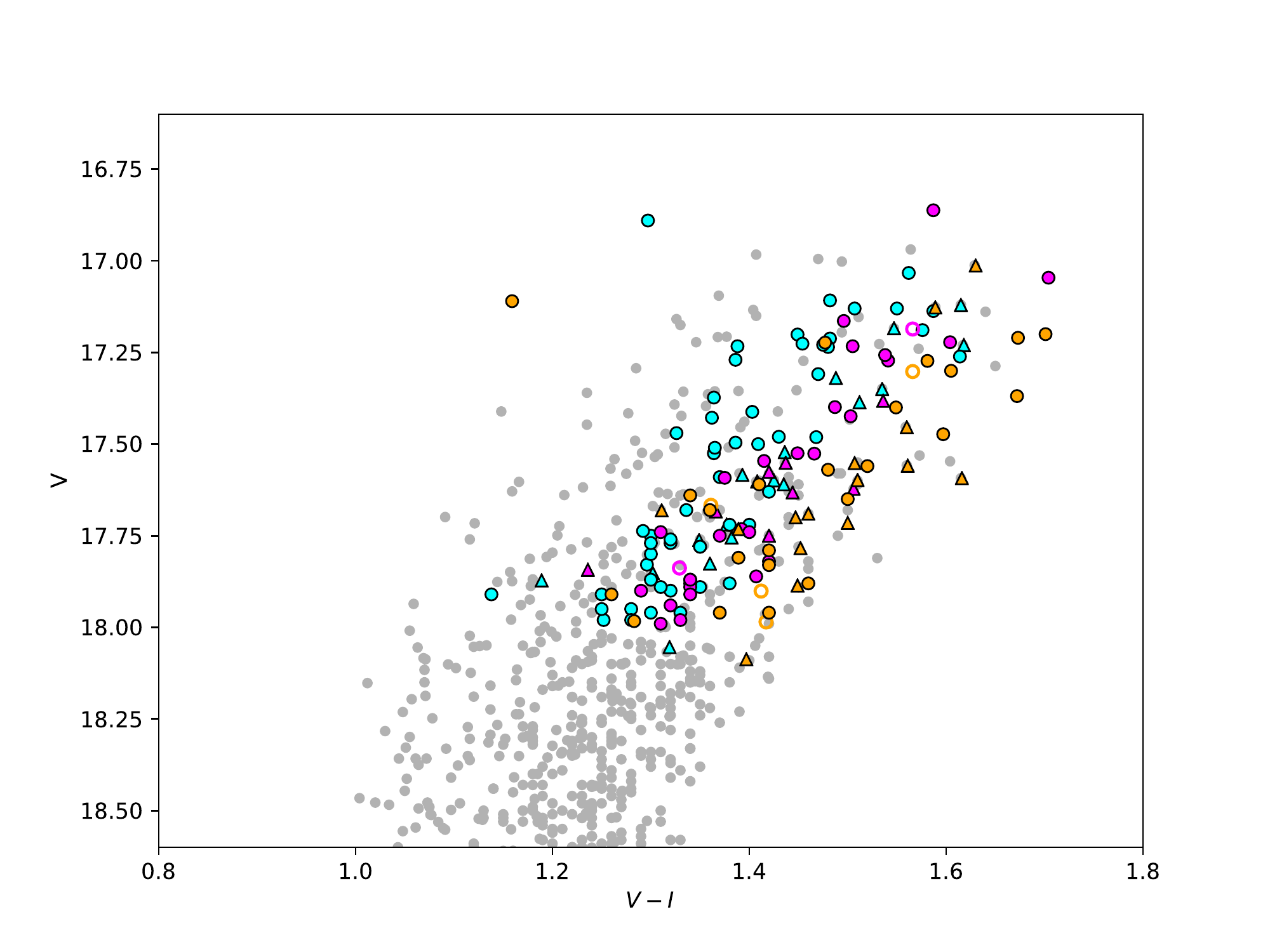}
\caption{Position of the target stars in the Sculptor red giant branch CMD\@. The observed stars are shown as triangles 
for the 45 stars observed with GMOS-S and as circles for the remaining 108 stars with red spectra only. Open markers indicate 
the 8 stars without a useable spectrum at the NaD lines while the grey small dots are for other Sculptor member stars from the studies of \citet{walker2009stellar}, \citet{coleman2005absence} and \citet{kirby2010multi}.  The photometry is taken solely from \citet{coleman2005absence}.  For the observed stars the colour coding refers to different metallicity groups: cyan, magenta and orange are used for stars
with [Fe/H] $<$ --1.9, --1.9 $\leq$ [Fe/H] $\leq$ --1.6 and [Fe/H] $>$ --1.6 respectively.}
\label{cmd}
\end{center}
\end{figure}

Our input data set consists of 161 Sculptor red giant stars that have been identified as Scl members by \citet{walker2009stellar}, \citet{coleman2005absence} and  \citet{battaglia2008analysis}. There are 114 stars brighter than $V=17.8$ and 47 stars with magnitudes $17.8 < V \leq 18.1$. The position of the targets on the Scl RGB is presented in the CMD shown in Fig. \ref{cmd}. The full sample covers an area in excess of $1^\circ$ in size, but the majority are found in a region $\sim$20' in diameter. 
The original intention was to observe the full sample with the Anglo-Australian Telescope (AAT) at Siding Spring Observatory using the 2dF multi-object fibre positioner and the AAOmega dual beam spectrograph \citep{saunders2004aaomega, sharp2006performance}.  The spectrograph was configured with the 
1700B grating in the blue arm, which provides a resolution of 1.1 \AA \, and the 2000R grating in the red arm, which provides a spectral resolution of 0.8 \AA.  With this configuration, the observations cover the G-band (CH, $\sim$4300\AA) and the violet $\sim$3880\AA\/ and blue $\sim$4215\AA\/ CN-bands in the blue spectra, and the NaD lines in the red spectra whose coverage is approximately 5800--6255 \AA.  

Unfortunately, despite multiple night allocations across two successive observing seasons, the conditions did not permit usable blue spectra to be obtained for the Sculptor red giants.  Nevertheless, good blue spectra were obtained for red giant stars in six globular clusters (M30, M55, NGC~2298, NGC~6752, NGC~288 and NGC~1851), and for Sculptor and GGCs red giants at the red wavelengths. The red spectra for the Sculptor stars consist of two distinct sets of combined multiple 1800 second exposures collected across 3 nights in October 2011 that resulted in 153 usable spectra at the NaD wavelengths. The red and blue spectra of the GGC stars were obtained in September 2010 and October 2011.  This data set consists of blue and red spectra for 196 stars belonging to the six GGCs, however, after the rejection of stars identified as likely AGB stars in photometric catalogues for the clusters, and removal of low signal-to-noise spectra, 164 GGC red giant spectra remain.
Because of the lack of success with the AAT blue spectra, a subset of 45 Sculptor red giants was subsequently observed with the Gemini-South telescope (program GS-2012B-Q-5) and the Gemini Multi-Object Spectrograph (GMOS-S; \citep{hook2004gemini}). The B1200 grating was used with 1 arcsec slits to yield a resolution of 2.3\AA\/.  Six masks were observed with two integrations per mask at slightly different central wavelengths in order to compensate for the inter-chip gaps.
The GMOS-S spectra were reduced, extracted and wavelength calibrated with standard IRAF/Gemini software\footnote{\small{IRAF is distributed by the National Optical Astronomy Observatories, which are operated by the Association of Universities for Research in Astronomy, Inc.,  under cooperative agreement with the National Science Foundation.}}. The red spectra for the Sculptor stars and the blue and red spectra for the GGC giants were reduced with the 2dF data reduction pipeline 2dfdr\footnote{\small{https://www.aao.gov.au/science/software/2dfdr}}. 
Details of the observations are shown in Table 1.

In order to provide additional confirmation of the membership in the Sculptor dSph galaxy of the stars observed in this study, we have investigated the individual proper motions ($\mu_{RA}$,  $\mu_{Dec}$) available from the Gaia DR2 data release \citep{gaia2018}.  We find that all of the stars fall inside a circle centred on the mean values of $\mu_{RA}$ and $\mu_{Dec}$ with radius equal to 3$\sigma$, where $\sigma$ is the standard deviation of the radial distances from the centre.  We have therefore no reason to doubt the Scl membership of any star in this sample.

\begin{table}
\caption{Details of the observations.}
\resizebox{\linewidth}{!}
{\begin{threeparttable}
\renewcommand{\TPTminimum}{\linewidth}
\centering 
\begin{tabular}{l c c c r} 
\hline 
\hline
\normalsize{} & \normalsize{Blue (GMOS-S) } & \normalsize{Blue (AAOmega)} & \normalsize{Red 	(AAOmega)} & \\ [0.5ex] 
\normalsize{} & \normalsize{Sculptor } & \normalsize{GGCs} & \normalsize{Sculptor and GGCs} & \\ [0.5ex] 
\hline 

Grating				&	B 1200	& 1700B	& 2000R 						&\\
Resolution ($\AA$)	&	2.3		& 1.1	& 0.8 							&\\
Observed stars		&	45		& 196	& 153\tnote{b} \, - 196\tnote{c}	&\\
Central Wavelength 	&	4400 \AA \,and 4350 \AA\tnote{a}	& 4100 \AA & 6050 \AA &\\

\hline 

\end{tabular}

\begin{tablenotes}
\item [a] Observed at two different central wavelengths
\item [b] Useful Sculptor spectra
\item [c] Globular Cluster spectra. 164 GGC spectra were used after excluding AGB and low S/N spectra.
\end{tablenotes}
\label{Tabla_obs} 

\
\end{threeparttable}}
\end{table}

\subsection{Feature strengths}
\label{strengths}

The analysis of CH- and CN-bands was based on the measurement via numerical integration of three indices that are sensitive to the strength of the bands in the spectra.  In what follows $F_{\lambda}$ is the intensity, and $\lambda$ the wavelength, both from (pseudo-) continuum-normalized spectra. The spectra were (pseudo-) continuum-normalized by using standard IRAF routines: a low-order polynomial function was fit to the stellar (pseudo-) continua in order to consistently remove the overall shape of the spectra imposed by the convolution of the instrument response and the spectral energy distributions of the stars.  The spectra were then shifted to rest wavelength using the velocity derived from each observed spectrum.  All subsequent analysis uses the (pseudo-) continuum-normalized, velocity-corrected spectra.

We first measured the strength of the $\lambda$3883 \AA \, CN-band by generating the index $S(3839)$ \citep{norris1981abundance}, which compares the intensity within the violet CN-band with that of the nearby continuum.  

\begin{equation}
S(3839) = - 2.5 \, log_{10} \int_{3826}^{3883} F_{\lambda} d\lambda \,\,\, \,  \textit{{\LARGE /}} \, \int_{3883}^{3916} F_{\lambda} d\lambda
\label{eq_S3839}
\end{equation}

Second, we have measured the strength of the cyanogen band at $\lambda$4215 \AA\/ via the index $S(4142)$, which is explained in detail in \citet{norris1979cyanogen}:

\begin{equation}
S(4142) = - 2.5 \, log_{10} \int_{4120}^{4216} F_{\lambda} d\lambda \,\,\, \,  \textit{{\LARGE /}} \, \int_{4216}^{4290} F_{\lambda} d\lambda
\label{eq_S4142}
\end{equation}
Finally, we measured the strength of G-band at $\sim$4300 \AA\/ by employing the $W(G)$ index given by \citet{norris1984anticorrelation}: 

\begin{equation}
W(G) = \int_{4290}^{4318} \left(   1 -  F_{\lambda} / F_{4318}  \right)   d\lambda \, , 
\label{eq_wg}
\end{equation}
where $F_{4318}$ comes from the mean of 5 maximal intensities in the range of wavelengths $\lambda\lambda$ 4314 -- 4322 \AA. 

As noted above, we employed the (pseudo-) continuum-normalized spectra for the band strength measurements; we have not attempted to flux-correct the spectra, a process that can be problematical for multi-fibre observations.  Consequently, it is not possible to directly compare our band strength measurements with those made on flux-corrected spectra.  

\begin{figure*}
\begin{center}
\includegraphics[width=18cm]{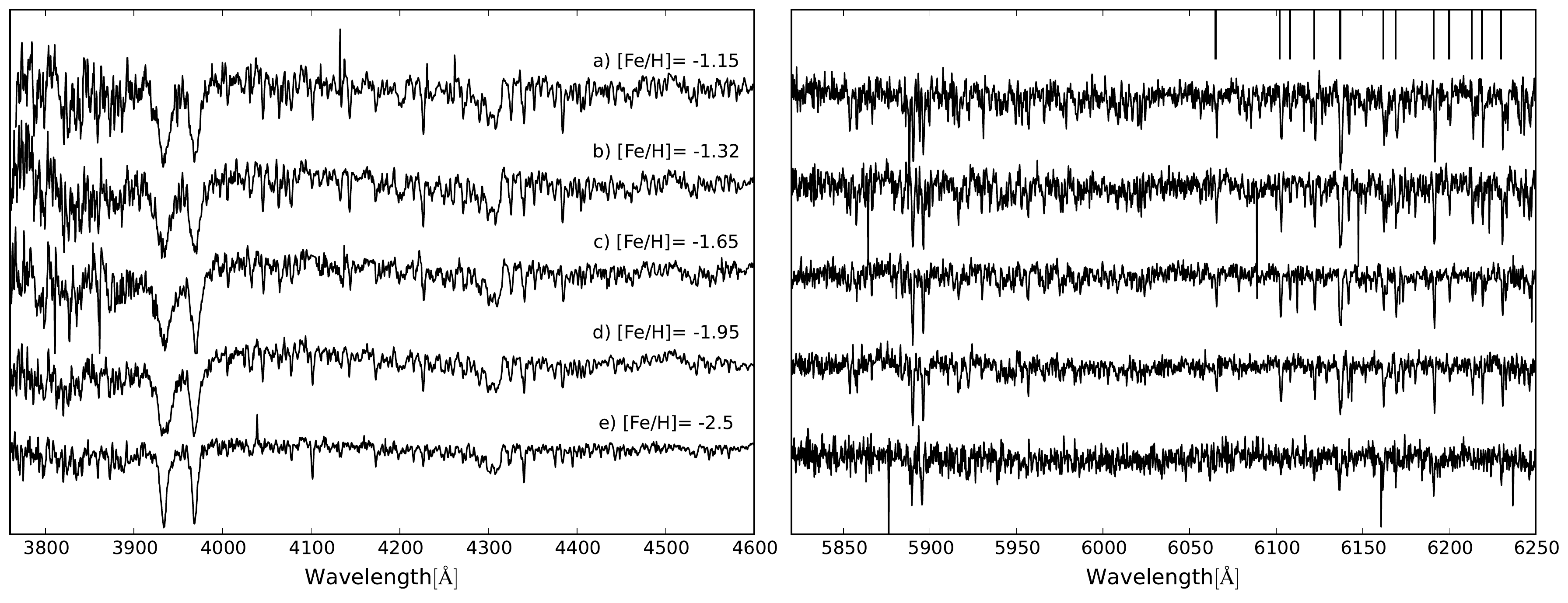}
\caption{Examples of the continuum-normalized velocity-corrected spectra of Scl target stars at different metallicities. The spectra have been shifted vertically and the metallicity values are indicated.  The left panel shows GMOS-S Gemini blue spectra, while red spectra from AAOmega are shown in the right panel. The colours and magnitudes of the stars are as follows:
a) Scl-0233: $V$=17.55, $V-I$=1.50; b) Scl-0470: $V$=17.78, $V-I$=1.45; c) Scl23: $V$=17.84, $V-I$=1.23; d) Scl-0468: $V$=17.61, $V-I$=1.43; e) Scl-0272: $V$=17.58, $V-I$=1.39. The vertical lines in the upper portion of the right panel indicate the location of the spectral lines used to determine the metallicities.} 
\label{blue_red_spectra}
\end{center}
\end{figure*}

The CH and CN indices were measured on the Sculptor GMOS-S spectra, which have a resolution of 2.3 \AA, and on the GGC AAOmega spectra that have a resolution of 1.1 \AA. In order to check the effect of the different resolutions on the measurements, we smoothed the blue spectra of a selection of GGCs stars to match the resolution of the GMOS spectra, and remeasured the indices. We found that, on average, the difference for $S(3839)$ is only 0.01 mag (smaller at lower resolution). 
Similarly, we tested the effect of changing resolution on the $W(G)$ values. On average, the difference in $W(G)$ is $\sim$1.3 \AA, which in the context of the average of $W(G)$ measurements, corresponds 
to $\sim$13\% decrease at the lower resolution.  We have not made any adjustments for this effect.
Finally, the equivalent width (EW) of the Na~D sodium absorption lines at $\sim$5889\AA\, and $\sim$5895 \AA, were determined via Gaussian fits, using standard routines of IRAF, and the continuum-normalized velocity-corrected spectra.  
Examples of the blue (GMOS) and red (AAOmega) spectra for Sculptor stars with different metallicities are shown in Fig. \ref{blue_red_spectra}.

\section{Metallicities}
\label{metallicities_procedure}

Because Sculptor has a wide range of metallicities we analysed the CH- and CN-band strengths for three distinct groups of Scl members: [Fe/H]$\leq$--1.9 (metal-poor stars), --1.9$<$[Fe/H]$<$--1.6 (intermediate-metallicity stars) and [Fe/H]$\geq$--1.6 (metal-rich stars), as shown in Fig. \ref{wg_cn_metall}. To determine the metallicities of the Sculptor stars we followed an approach similar to that described in, e.g., \citet{norris1983oxygen}, \citet{frebel2007chemical}, and \citet{norris2012oxygen}.  In particular, we selected 12 strong lines of calcium, iron and nickel in the RGB-star red spectra for use in the metallicity determination.  The lines used are $\sim$6102 \AA\,, 6122 \AA\,, 6162 \AA\,and 6169 \AA\, for Ca\thinspace I lines; $\sim$6065 \AA\,, 6137 \AA\,, 6191 \AA\,, 6200 \AA\,, 6213 \AA\,, 6219 \AA\, and 6230 \AA\, for Fe\thinspace I lines; and $\sim$6108 \AA\, for Ni\thinspace I.  A small region around each line was extracted from the continuum-normalized velocity-corrected spectra and the wavelength scale offset by the rest wavelength (in air) of the line, so that the line centre was at zero wavelength. 
The wavelength-offset regions for each line were then combined using standard IRAF routines to form a single line profile.  This process has the effect of increasing the signal-to-noise in the combined line profile facilitating the measurement, via fitting a Gaussian profile, of the equivalent width.  The approach was applied to the red spectra of the GGC stars as well as to the red spectra of the Scl members.

\begin{figure}
\begin{center}
\includegraphics[width=8.5cm]{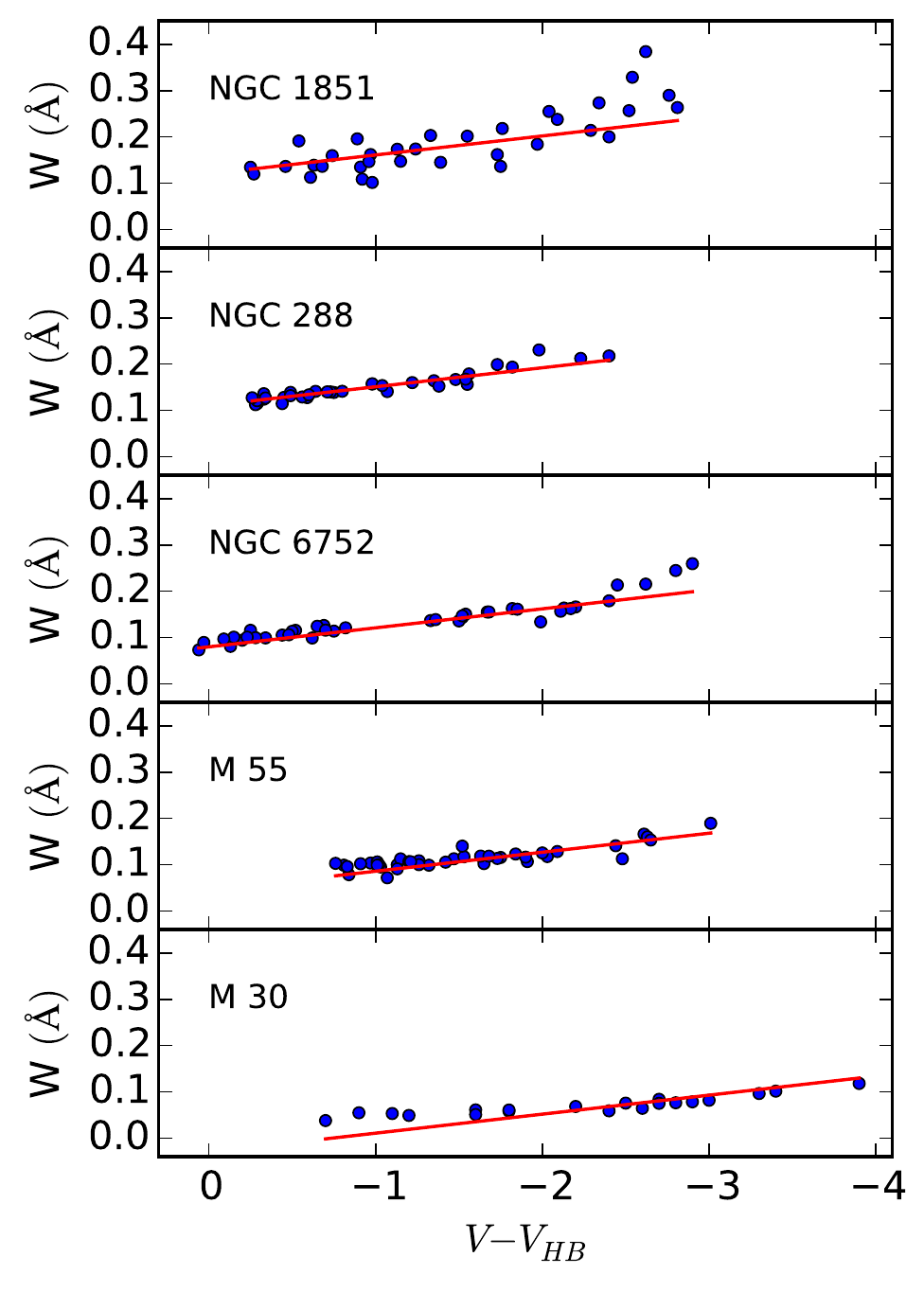}
\caption{Plot of the equivalent width of the combined Ca, Fe and Ni lines against $(V-V_{HB})$ for the 5 GGC calibration clusters. The clusters are ordered in decreasing metallicity from top to bottom. The red line is the mean slope, $\alpha = -0.04\, \AA\, mag^{-1}$.}
\label{calibration_cluster}
\end{center}
\end{figure}

In Fig.\ \ref{calibration_cluster} we show the resulting EW values for the cluster stars plotted against $V-V_{HB}$ where the $V_{HB}$ values were taken from the current on-line version of the \citet{harris1996catalog} catalogue (see Table \ref{tabla_harris}). For each cluster the line of best-fit was calculated, and, since the slopes did not vary significantly from cluster to cluster, the average slope was determined and refit to the cluster values, as shown in Fig.\ \ref{calibration_cluster}. The slope is  $\alpha$ = --0.04 \AA\,mag$^{-1}$ and the error in the slope is $\pm 0.013$, the standard error of the mean of the slope estimates for the five GGCs. 
The error in the slope will be propagated into the calculation of the reduced equivalent width (defined in the next paragraph). 
We note that the GGC NGC~2298 has not been considered in this calibration process because the small number of stars in this cluster with sufficient S/N yielded a slope that was substantially different than that for the other clusters.

\begin{figure}
\begin{center}
\includegraphics[width=8.5cm]{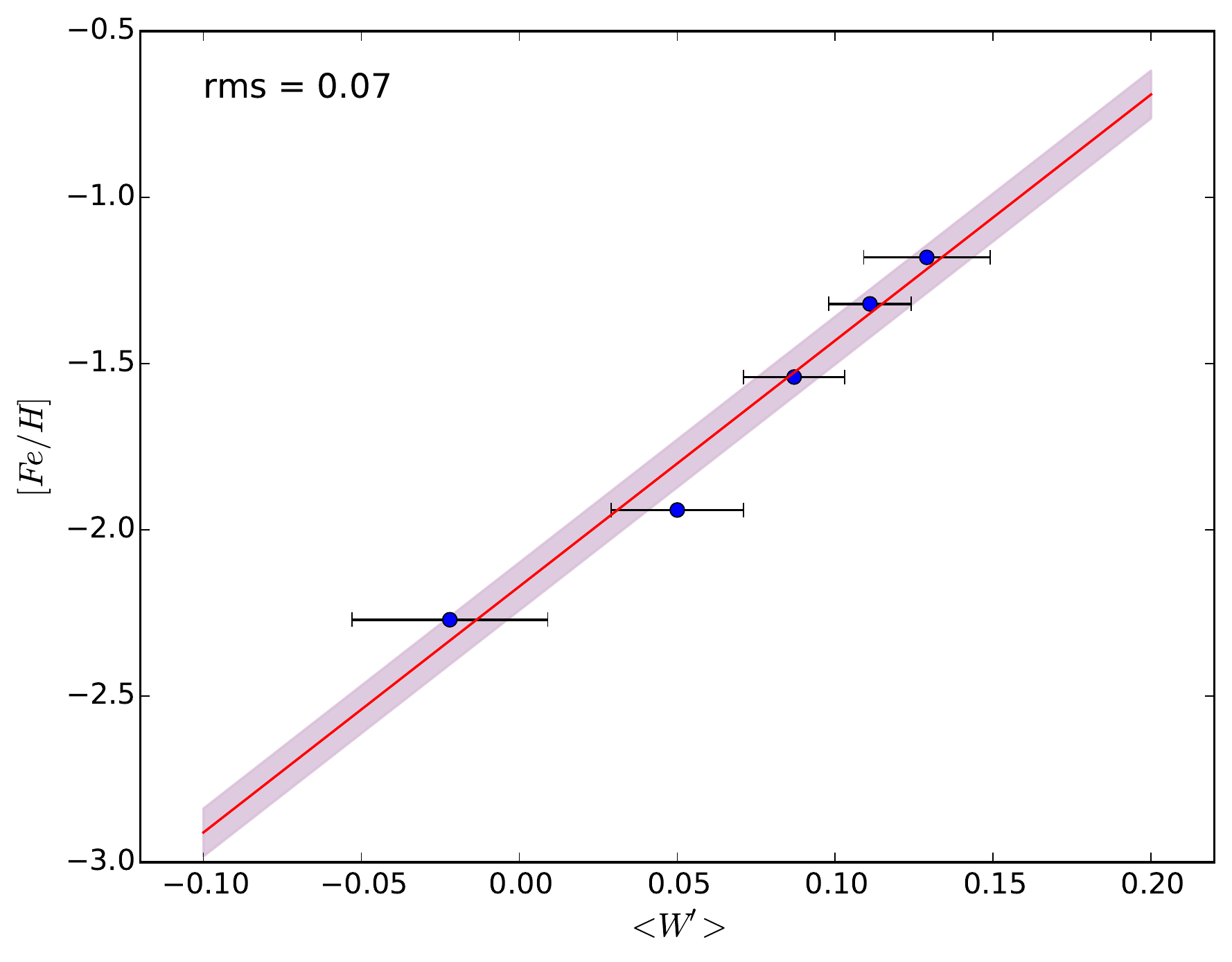}
\caption{[Fe/H] against <W'> for the 5 calibration globular clusters. The metallicities  were taken from the most recent on-line version of the \citet{harris1996catalog} catalogue (see Table \ref{tabla_harris}). The red line is the best fit to the data and the shaded area is the rms dispersion.  The error bars shown are the  mean error calculated from the propagation of the uncertainties for each set of cluster $W'$ values.} 
\label{calibration_fe}
\end{center}
\end{figure}

We then used $\alpha$ to set the reduced equivalent width $W'$ for each star, with $W'$ defined by $W' = (EW) - \alpha (V-V_{HB})$.  The average reduced equivalent width for each GGC was then computed and plotted against the cluster metallicities from the current on-line version of the \citet{harris1996catalog} catalogue (see Table \ref{tabla_harris}). The error bars correspond to the mean error calculated from
the propagation of the uncertainties in quadrature and then divided
by the number of values for each set of cluster $W'$ values. The result is shown in Fig. \ref{calibration_fe} together with a linear fit to the data points.   The relation found  is $[Fe/H] = 7.40 \pm 0.18$<$W'$> $-2.17 \pm 0.18$ and the rms dispersion about the fit is 0.07 dex. We then used this relation to compute the metallicity for each star in our Scl sample from the measured EW values and each star's $V-V_{HB}$ value, using a value of $V_{HB}$ = 20.35  for Scl \citep{tolstoy2001using}.  There are two sets of Scl spectra so the two EW measurements were averaged to determine the final value.  The mean difference between the two sets of $W'$ then enables us to estimate a typical metallicity error: the standard deviation of the differences is of order 0.01 -- 0.015 \AA\/ ($\sim$10--15\% of a typical EW value of 0.1\AA) and with the calibration this corresponds to a metallicity error of $\sim$0.1 dex.

\begin{table}
\caption{Adopted parameters from \citet{harris1996catalog} catalogue}
\resizebox{\linewidth}{!}
{\begin{threeparttable}
\renewcommand{\TPTminimum}{\linewidth}
\centering 
\begin{adjustbox}{max width=8.5cm}
\begin{tabular}{l c c c c r} 
\hline 
\hline
\normalsize{GGC ID} & \normalsize{[Fe/H]} & \normalsize{V$_{HB}$} & \normalsize{(m-M)$_{V}$} & \normalsize{E(B-V)} & \\ [0.5ex] 
\hline 

M~30 		&	-2.27 & 15.10 & 14.64 & 0.03 &\\
M~55			&	-1.94 & 14.40 & 13.89 & 0.08 	&\\
NGC~6752		&	-1.54 & 13.70 & 13.13 & 0.04	&\\
NGC~288		& 	-1.32 & 15.44 & 14.84 & 0.03 &\\
NGC~1851		& 	-1.18 & 16.09 & 15.47 & 0.02 &\\
\hline 

\end{tabular}
\end{adjustbox}
\end{threeparttable}}
\label{tabla_harris} 
\end{table}

\begin{figure}
\begin{center}
\includegraphics[width=8.5cm]{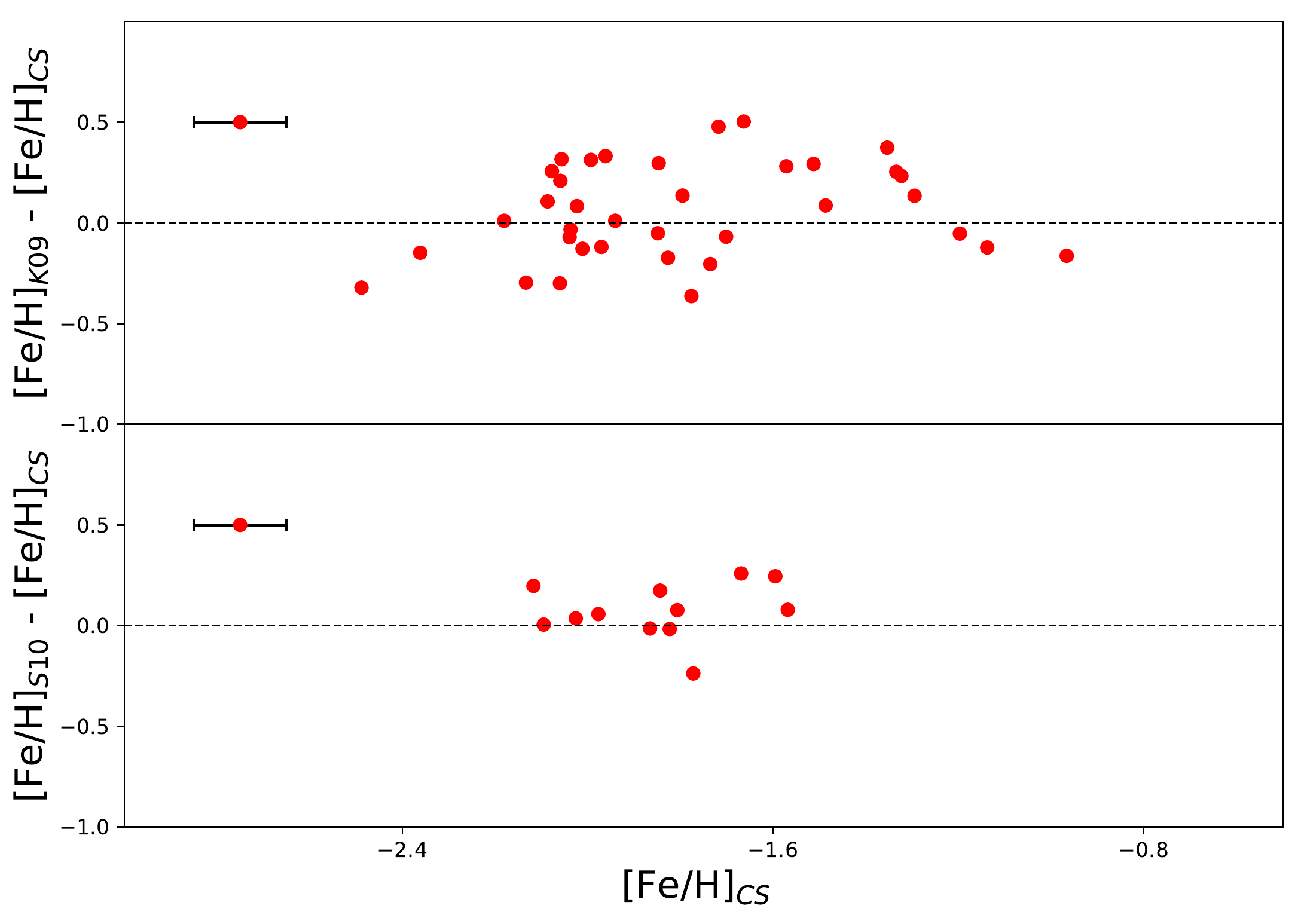}
\caption{Estimation of [Fe/H] for Sculptor stars compared with literature values. Upper panel shows the difference between [Fe/H] found in \citet{kirby2009multi} and [Fe/H] from the current calibration ([Fe/H]$_{CS}$) against [Fe/H]$_{CS}$. Lower panel shows the difference between \citet{starkenburg2010nir} and  [Fe/H]$_{CS}$ against [Fe/H]$_{CS}$.  Error bars ($\pm$1$\sigma$ with $\sigma$=0.1 dex) for the [Fe/H]$_{CS}$ values are shown.}
\label{metallicities}
\end{center}
\end{figure}

We have then compared our Scl metallicity values with the literature results of \citet{kirby2009multi} and \citet{starkenburg2010nir}. First, our determination of the mean metallicity for Scl, based on the 146 stars for which we have a metallicity estimate, is $[Fe/H]_{mean}$ = --1.81 dex.  This value is entirely consistent with the mean metallicity  found by \citet{kirby2009multi}, $[Fe/H]_{mean}$ = --1.73 dex, and that given by  \citet {starkenburg2010nir}, $[Fe/H]_{mean}$ = --1.77 dex.  Second, for the stars in common, we show in the panels of the Fig. \ref{metallicities} the difference between the abundances in \citet{kirby2009multi} and our abundances (upper panel), and the difference between the abundances in \citet{starkenburg2010nir} and our values (lower panel) plotted against our determinations. 
In each case there is no indication of any systematic offset in the abundance differences and the standard deviation of the differences are 0.23 dex (upper panel) and 0.15 dex (lower panel) respectively. 

\citet{kirby2009multi} indicates that a typical error in their metallicity determinations is of order $\varepsilon$ = 0.10 dex, while for \citet{starkenburg2010nir} the listed typical uncertainty is $\varepsilon$ = 0.13 dex.  If we assume these typical errors are valid, then we can use them and the dispersions in the abundance differences to provide an alternative estimate of the uncertainty in our abundance determinations.  From the comparison with the \citet{starkenburg2010nir} values we derive an estimate of 0.07 dex for our abundance uncertainties, which is quite consistent with that derived above.  The comparison with the \citet{kirby2009multi} values, however, yields a substantially larger estimate (0.21 dex) for the uncertainty in our abundance determinations that seems at odds with the comparison with the \citet{starkenburg2010nir} values and with our internal error estimate.  It is possible that \citet{kirby2009multi} have underestimated their abundance errors.  We therefore adopt a typical abundance error of 0.10 dex for our determinations.

Given the metallicity spread in Scl, there is no straightforward way to decide if a particular Scl star lies on the RGB or on the asymptotic giant branch (AGB) in the CMD\@.  At fixed luminosity, an AGB star is hotter than an RGB star of the same metallicity leading to weaker lines in the AGB-star spectrum.  Consequently, our abundance determination process will underestimate the actual abundance of Scl AGB stars.  We can estimate the size of this effect by using our GGC star spectra, noting that the GGCs used are mono-metallic.  Specifically, in our observed sample for NGC~1851, there are 3 stars that are readily identified as AGB stars in the cluster CMD, while there are 4 such stars in the NGC~6752 sample. The mean [Fe/H] for the NGC 1851 AGB stars from our abundance analysis approach is [Fe/H] = --1.40 ($\sigma$ = 0.13), while for the NGC 6752 AGB stars the mean abundance is [Fe/H] = --1.75 ($\sigma$ = 0.02).  Both these values are about 0.2 dex less than the metallicities given in the  \citet{harris1996catalog} catalogue (see Table \ref{tabla_harris}).  In GGCs $\sim$20-25\% of giant branch stars belong to the AGB and we assume that fraction also applies in Scl.  Consequently, we must remain aware that our metallicity estimates for $\sim$ 20-25\% of the Scl sample could be underestimated by up to 0.2 dex.

\section{Results}
\label{results cn ch na}


\subsection{CH and CN}

The GGC abundance anomalies were originally characterised via analysis of
red giant spectra covering the CN-bands around 3880 \AA\/ and 4215 \AA, and the 
CH-band at $\sim$4300 \AA \, commonly known as the G-band
\citep[e.g.,][]{OsbornW1971,norris1981abundance,cannon1998carbon,carretta2005abundances}. 
The data revealed that the stars in a given cluster could classified into two well separated groups, one CN-strong and the other CN-weak.  There is also a well defined anti-correlation in that the CN-strong stars are relatively CH-weak and vice versa.  At the same time spectrum synthesis calculations revealed that the effects are driven by enhanced nitrogen and depleted carbon abundances in the CN-strong population relative to the CN-weak group \citep[e.g.,][]{pancino2010,kraft1994abundance}. 
To investigate if these GGC effects are also present in the Sculptor red giant data set we measured the indices $W(G)$, $S(3839)$ and $S(4142)$ as outlined in \ref{strengths} for the Scl stars with blue spectra.  As noted in Section \ref{metallicities_procedure}, we then separate the set of index measurements into three metallicity groups using the [Fe/H] values derived from the corresponding red spectra.  The IDs, positions, $V$ and $V-I$ photometry, index measures, sodium line strengths (see Section \ref{na_sect}) and derived metallicities for these stars are listed in Table \ref{tabla_total}. 

\begin{figure*}
\begin{center}
\includegraphics[width=16cm]{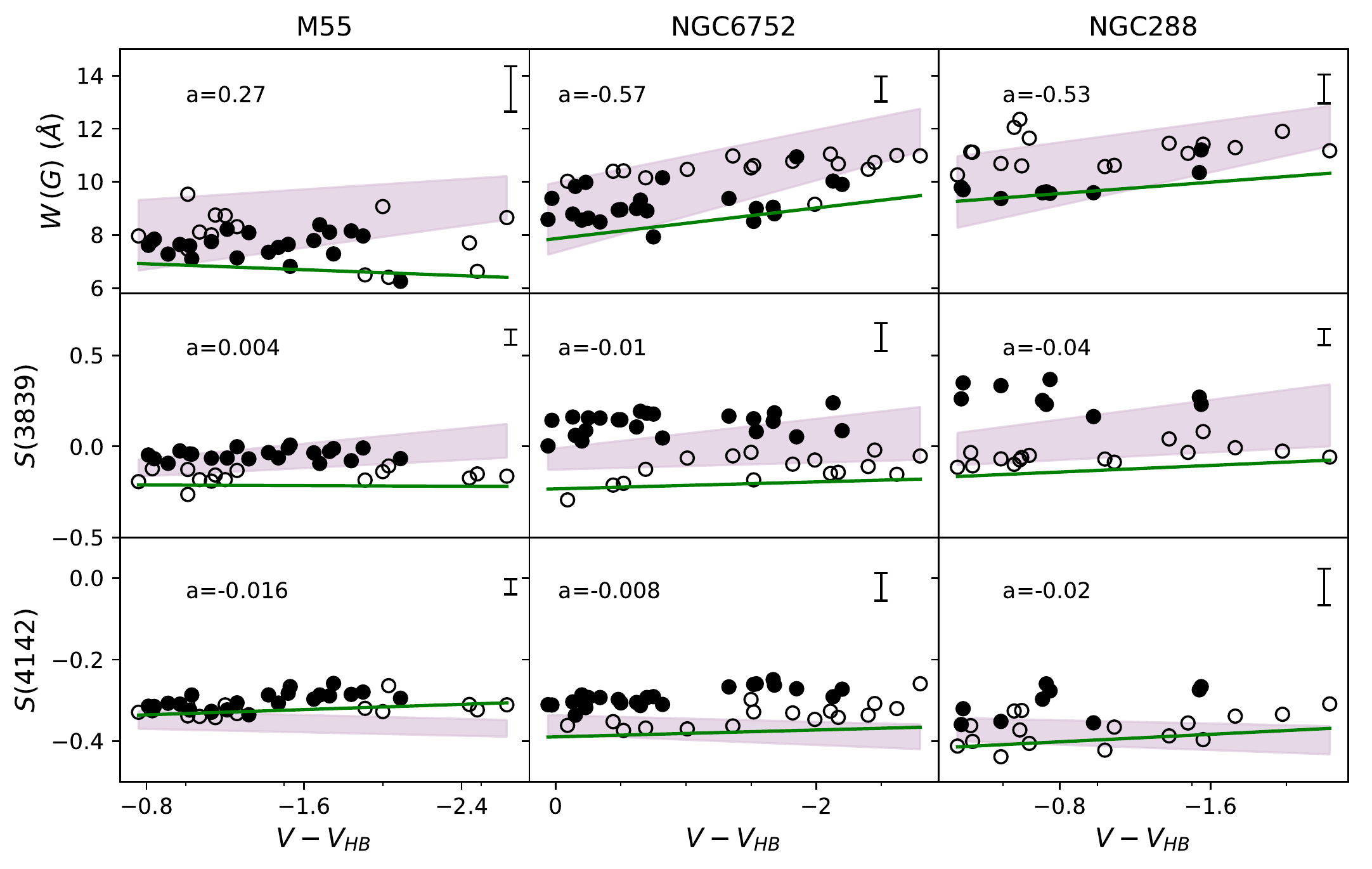}
\caption{Dependence of the band-strength indices $W(G)$, $S(3839)$ and $S(4142)$ on $V-V_{HB}$ for RGB members of the GGCs M55, NGC~6752 and NGC~288. The central row is used to define CN-strong and CN-weak stars, which are represented by filled and open circles, respectively. The green straight lines are the adopted lower envelopes to the data for each index and cluster, and the corresponding slope values `a' are given on the panels. The measurement error associated with each index and cluster is plotted in the top right corner of each panel. The error bars shown are $\pm$1$\sigma$ in length.  The shaded area in each panel is explained in Section. \ref{spec_synth}.} 
\label{wg_cn_clusters}
\end{center}
\end{figure*}

\begin{figure*}
\begin{center}
\includegraphics[width=16cm]{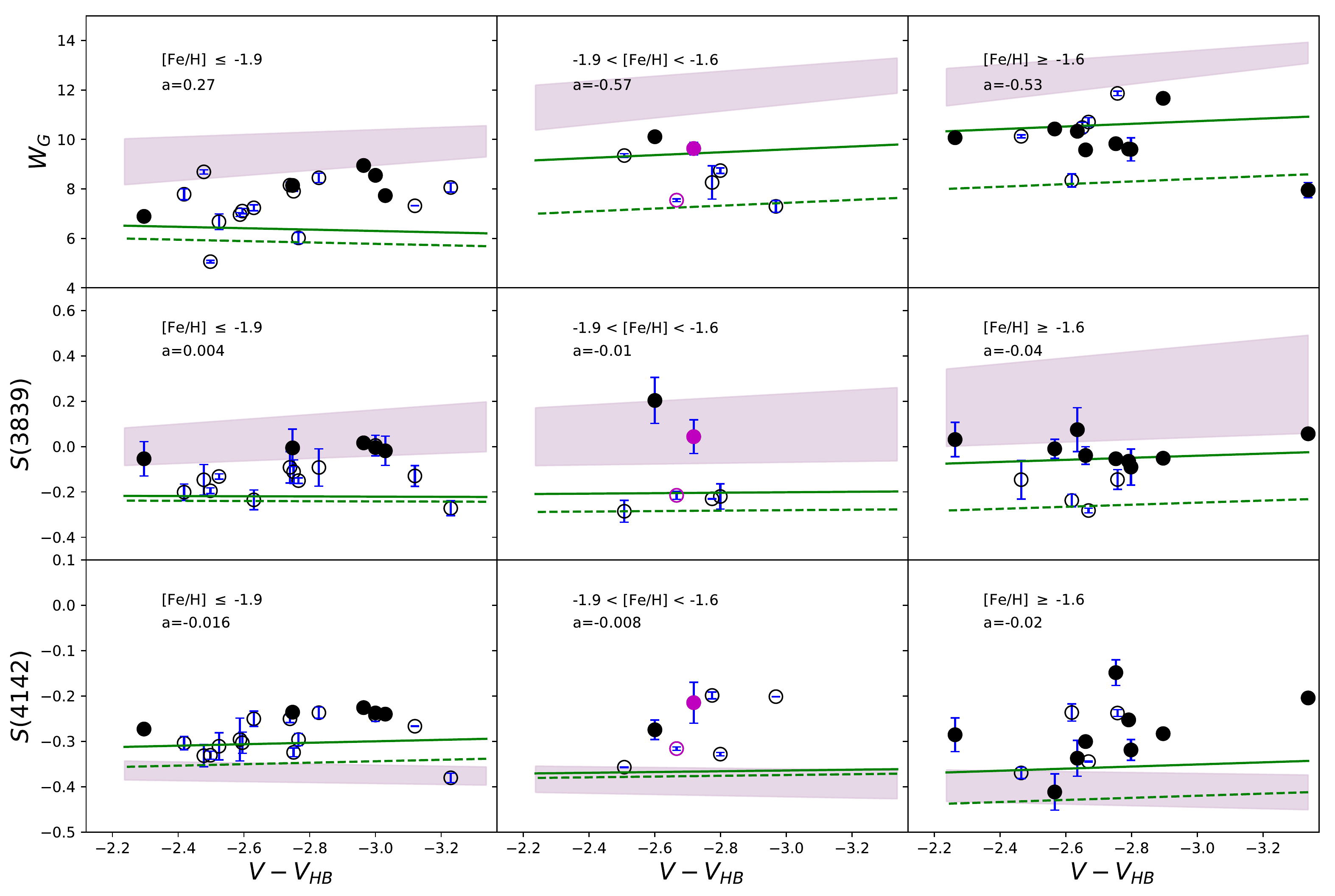}
\caption{Dependence of the band-strength indices $W(G)$, $S(3839)$ and $S(4142)$ on $V-V_{HB}$ for the Scl stars, grouped into three metallicity ranges as indicated at the top of each panel.  Stars characterised as CN-strong (from the $S(3839)$ index) are plotted as filled circles, open circles are CN-weak stars. The green dashed lines are the adopted lower envelopes to the Scl data. Their slopes (a), which are identical to those shown in Fig.\ \ref{wg_cn_clusters}, are shown.  The green solid lines are the lower envelope lines from the GGC sample (Fig.\  \ref{wg_cn_clusters})  which have been extrapolated to higher luminosities when necessary.  Individual uncertainties are shown for each star.  
The magenta open and filled circles (CN-weak and CN-strong, respectively) indicate stars  Scl-0492 and Scl-0247 that are discussed in Section \ref{synthCN}, and correspond to the stars in Fig.\ \ref{comp_scl}.  The shaded area in each panel is explained in Section. \ref{spec_synth}. } 
\label{wg_cn_metall}
\end{center}
\end{figure*}

For comparison purposes we also measured, in the same way as for the Scl stars, the band strength indices for red giants in three GGCs using the AAOmega blue spectra.   The three GGC chosen (M55, NGC~6752 and NGC~288) have metallicities that are comparable to the mean metallicities of the three Scl groups.  The resulting GGC band strengths, plotted against $V-V_{HB}$, are shown in Fig.\ \ref{wg_cn_clusters}.   We use the values of the $S(3839)$ index (middle panels) to classify the GGC stars as CN-strong or CN-weak via a visual inspection of the distributions.  As expected, there is a clear separation between the two groups.  The figure also shows a green straight line in each panel that represents the lower envelope of the data for each index and cluster. These lines represent an assumed minimum value for the indices at any V - VHB. The slope of these lines was obtained from the best fit  for all the stars in each panel and the vertical location was set to encompass the minimum  values, given the index errors.  Based on these lines we have defined a parameter $\delta$, similar to the one introduced by \citet{norris1981abundance}, that measures the index displacement at a given $V$ magnitude with respect to the lower envelope. This excess parameter $\delta$ minimizes the effects on the band strength indices of the changing temperatures and surface gravities of the stars. \\
Nevertheless, each Scl metallicity group does contain a range in abundance and, at fixed $V-V_{HB}$, the higher metallicity stars in each group will have lower $T_{eff}$, and vice-versa, and this can potentially affect the range of line strength indices present.  We have explored the consequences of this effect by measuring the indices on a number of synthetic spectra that have the same resolution as the observed data.  Specifically, we note for each Scl metallicity group, the variation in $V-I$ at fixed $V-V_{HB}$ is typically $\pm$0.2, $\pm$0.17 and $\pm$0.19 mag for the metal-poor, intermediate-metallicity and metal-rich groups, respectively (see Fig. \ref{cmd}), presumably driven primarily by the metallicity ranges.  These colour variations were obtained via low-order polynomial fits to the giant branch photometry.  Then using the $T_{eff}$:V-I:[Fe/H] calibrations from \citet{ramirez2005effective},  the mean metallicity for each group, and a typical $V-I$ value, the colour variations correspond the temperature variations of $\sim\Delta$ 240 K, $\sim\Delta$ 210 K, and $\sim\Delta$ 240 K, respectively. We then calculated for each metallicity group the $W(G)$, S(3839) and S(4142) indices from synthetic spectra assuming the mean metallicity of each group with upper and lower temperature values that differ from the adopted values by these offsets, assuming the mean metallicity of each group. For the most metal-poor group, the resulting changes were $\sim$0.85 \AA\/ for $W(G)$, 
$\sim$0.024 mag for S(3839) and $\sim$0.017 mag for S(4142), respectively (higher values for lower temperatures).  For the 
intermediate-metallicity and metal-rich groups the changes were $\sim$0.69 \AA\, $\sim$0.029 mag and $\sim$0.017 mag, and 
$\sim$0.60 \AA, $\sim$0.027 mag,  and $\sim$ 0.017 mag, respectively.  Comparison with the panels of Fig.\ \ref{wg_cn_metall}
shows that these changes are substantially smaller than the observed index ranges for each metallicity group, indicating that the range in metalllicity (and thus in $T_{eff}$ at fixed $V-V_{HB}$) within each grouping is not a major contributor to the observed range in index values.

\begin{figure}
\begin{center}
\includegraphics[width=8.5cm]{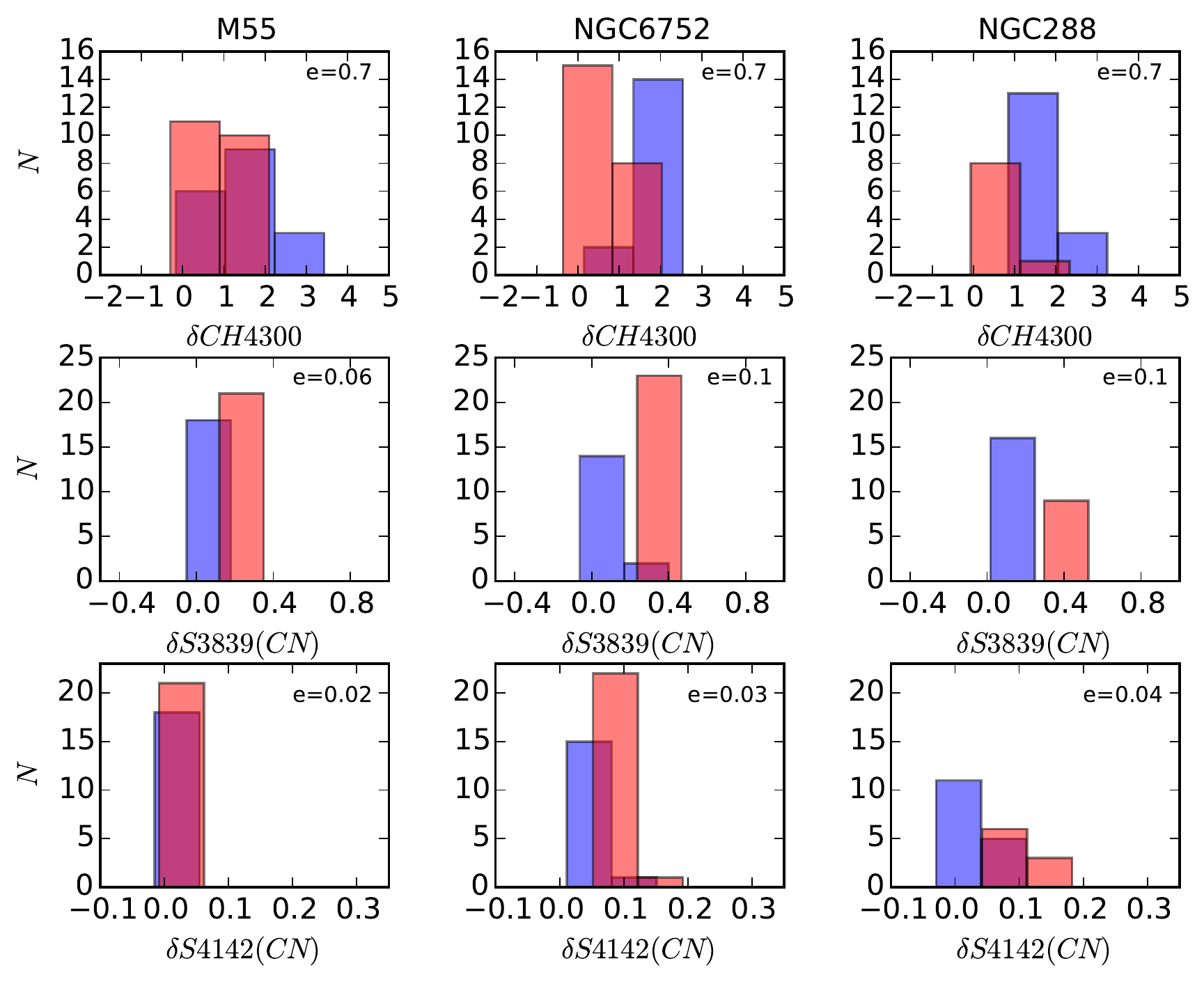}
\caption{Histograms of $\delta W(G)$ (top row), $\delta S3839$ (middle row) and $\delta S4142$ (bottom row) for the GGC data. The cluster names are indicated at the top of the columns. The bin sizes chosen are slightly larger than the index errors, which are given in top-right of each panel. CN-weak stars are shown in blue and CN-strong stars in red, with the classification coming from the middle row of Fig.\ \ref{wg_cn_clusters}.} 
\label{hist2} 
\end{center}
\end{figure}

Fig. \ref{hist2} then shows histograms of the $\delta$ parameter for each index in the GGC sample.  The histogram of the $\delta S(3839)$ parameter demonstrates the well-established bimodality of the CN-band strength for these GGCs. 
The other panels in Fig.\ \ref{wg_cn_clusters} and Fig.\ \ref{hist2} show that the bimodality of the CN-band strength is also seen in the $S(4142)$ index, particularly for the two more metal-rich GGCs, and that there is a general anti-correlation between the CN- and CH-band strengths.  The errors in the indices have been estimated via the reasonable assumption that in the absence of observational errors, the CN-weak sequences in each cluster should show zero index scatter; hence the rms about a line fitted to the CN-weak sequences provides a measure of the index errors.  The sizes of the error estimates are indicated on the panels of Fig.\ \ref{wg_cn_clusters}.

\begin{figure}
\begin{center}
\includegraphics[width=8.5cm]{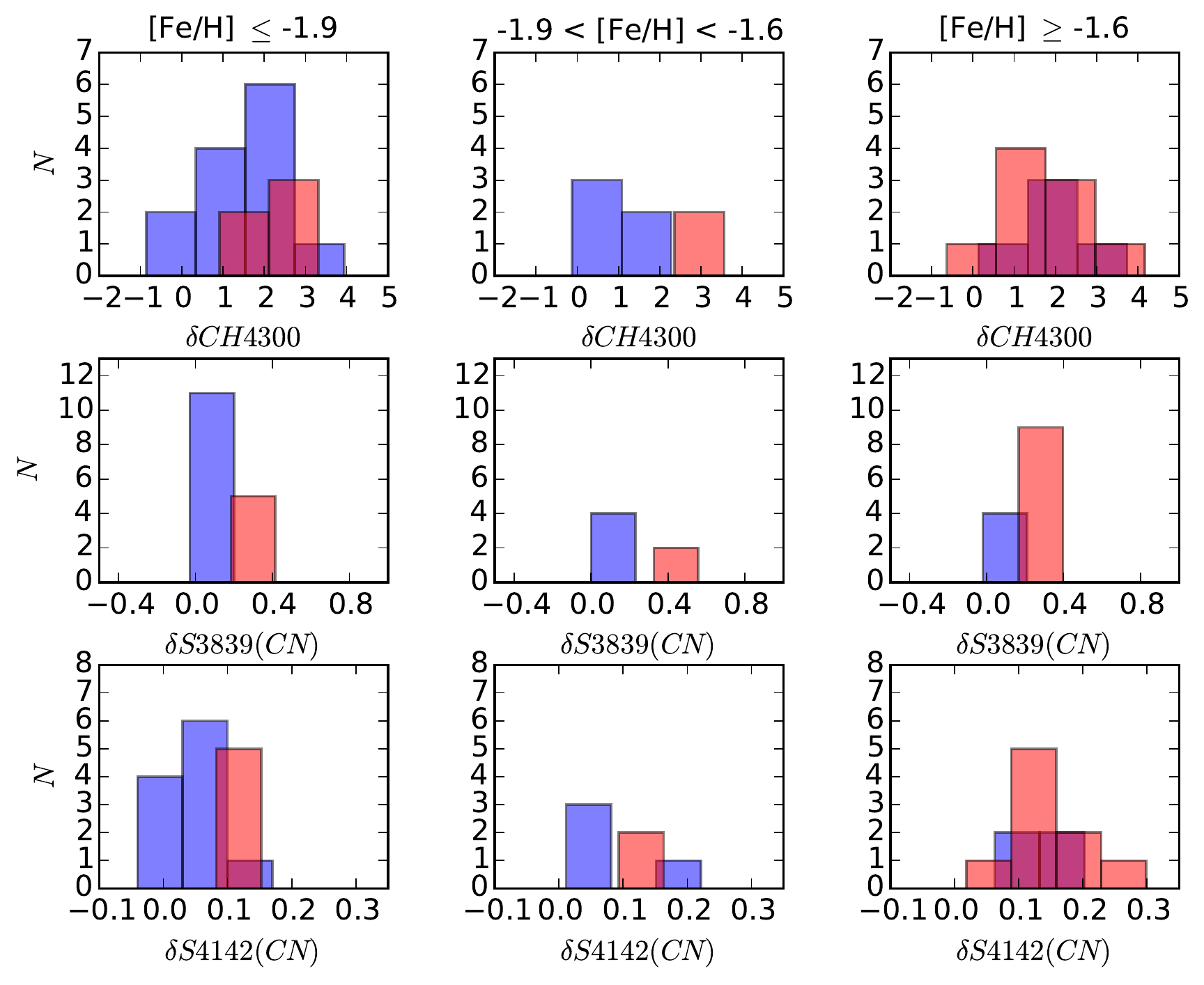}
\caption{Histograms of $\delta W(G)$ (top row), $\delta S3839$ (middle row) and $\delta S4142$ (bottom row) for the Scl stars. The Scl sample has been separated into three groups of metallicitiy as shown at the top of the columns.  The bin sizes chosen are slightly larger than the index errors in each group of metallicities. 
CN-weak stars are shown in blue and CN-strong stars in red, and were defined by the middle row of Fig.\ \ref{wg_cn_metall}.} 
\label{hist} 
\end{center}
\end{figure}

As for Fig.\ \ref{wg_cn_clusters}, Fig.\ \ref{wg_cn_metall} shows the band strength indices for the  Sculptor stars plotted against $V-V_{HB}$ magnitude, again using $V_{HB}$ = 20.35 for Scl \citep{tolstoy2001using}.  Shown also, as green dashed lines, are the adopted lower envelope lines for the Scl data where we have used the same slope as the corresponding lower envelope lines in the panels of Fig.\ \ref{wg_cn_clusters}.  These are also plotted in the Figure as green solid lines; in most of the cases there is a small offset between the line positions.  
As for the GGC data, we have measured the $\delta$ values for each index and metallicity group, and the histograms are shown in Fig.\ \ref{hist}.  We have then classified the Scl stars, by eye, as `CN-strong' or `CN-weak' on the basis of the $\delta S(3839)$ values.  The bimodality is not as clear-cut as it is for the GGC stars, at least in part because the index errors are larger for the Scl stars, which may lead to a blurring of any bimodality present.  In effect, it is the stars with the larger values of $\delta S(3839)$ that we have classified as CN-strong.  Only in the case of the intermediate metallicity group is there an 
indication of a gap in the $\delta S(3839)$ distribution that might indicate a bimodal distribution.
The errors in the index values for the Scl stars were estimated by making use of the fact that for each Scl star there are generally two measures of each index from the two GMOS-S spectra taken at slightly different central wavelengths.  The standard deviation of the differences (scaled by $\sqrt 2$) within each metallicity group was then employed as the index measurement uncertainty; these values are shown in the panels of Fig.\ \ref{wg_cn_metall}.  We note for completeness that we have not included in this analysis the star Scl-1013644, which has very strong CN- and CH-bands.  This star was identified as a CEMP-s star and has been discussed in \citet{salgado2016scl}.

In the panels of Fig. \ref{wg_cn_metall}, the lowest metallicity group shows good consistency between the $S(3839)$ and $S(4142)$ indices -- the stars classified as CN-strong stars in the $S(3839)$ panel also have larger values of $S(4142)$.  We have checked that this is not a metallicity effect -- the average metallicity for the CN-strong stars is this group is similar to that for CN-weak stars. However, the $W(G)$ index in this lowest metallicity group does not show any convincing indication that the CN-strong stars are significantly weaker in $W(G)$, as in seen in the comparison GGC M55.  Nevertheless, based on the mean $W(G)$ index error for this group, it does appear that there is a significant real dispersion in the $W(G)$ values.  

\begin{figure}
\begin{center}
\includegraphics[width=8.5cm]{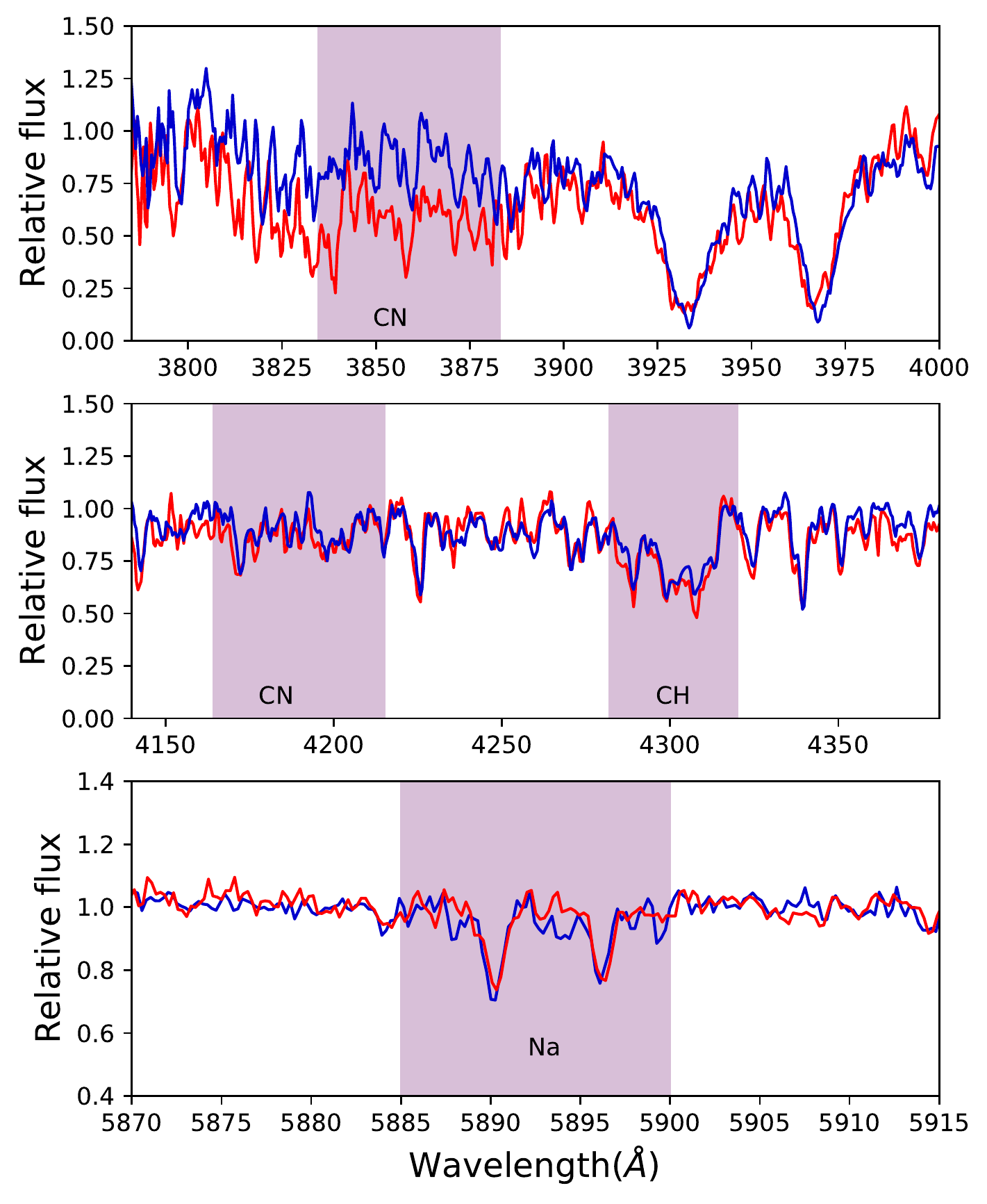}
\caption{Normalized spectra of Scl-0492 and Scl-0247, the stars highlighted in magenta in the central panel of Fig.\ \ref{wg_cn_metall}. The location of the CH, at $\lambda$ $\approx$ 4300\AA, and the CN, at $\lambda$ $\approx$ 3883\AA\/ and $\lambda$ $\approx$ 4215\AA, bands and Na, at $\lambda$ $\approx$ 5889\AA\/ and $\lambda$ $\approx$ 5895\AA, are shown by the shaded regions. The blue spectrum is Scl-0492 (CN-weak) while the red spectrum is Scl-0247 (CN-strong). 
} 
\label{comp_scl} 
\end{center}
\end{figure}

The same results are seen in the intermediate metallicity group -- the $S(4142)$ indices are broadly consistent with the classification based on the $S(3839)$ indices, but there is no indication that the Scl CN-strong stars are CH-weak in contrast to what is evident in the data for the corresponding GGC NGC~6752.  We illustrate this in the panels of Fig.\ \ref{comp_scl} where we show the continuum-normalised velocity-corrected observed spectra for the stars Scl-0247 (CN-strong) and Scl-0492 (CN-weak).  These stars are represented by the filled and open magenta circles in Fig.\ \ref{wg_cn_metall}.  The stars have similar colours and $V-V_{HB}$ magnitudes and thus similar $T_{eff}$ and $log~g$ values.   Our derived metallicities are also similar: we find [Fe/H] = --1.70 for Scl-0247 and [Fe/H] = --1.79 for Scl-0492.  The spectra shown in Fig.\ \ref{comp_scl} confirm the inferences from the indices in 
Fig.\ \ref{wg_cn_metall}: Scl-0247 has notably stronger 3880\AA\/ CN-strength but the differences at the 4215\AA\/ CN-band and, in particular, at the G-band (CH, $\sim$4300\AA) are much less substantial.  We defer to \S \ref{synthCN} a discussion of the C and N abundances that can be derived from these spectra via synthetic spectrum calculations.  Figure \ref{comp_scl} also shows, in the bottom panel, a comparison of the two spectra in the vicinity of the Na~D lines.  The  Na~D lines are clearly also very similar in strength, an outcome that will be discussed further in \S \ref{na_sect}.   

As for highest abundance group of Scl stars, the results are not as clear-cut.  The majority stars in this group are classified as CN-strong using the $S(3839)$ indices, but the CN-weak stars are not generally CN-weak in the $S(4142)$ panel.  Further, there is a broad range in the $W(G)$ values with no obvious separation of the CN-weak/CN-strong stars.

\begin{figure*}
\begin{center}
\includegraphics[width=16cm]{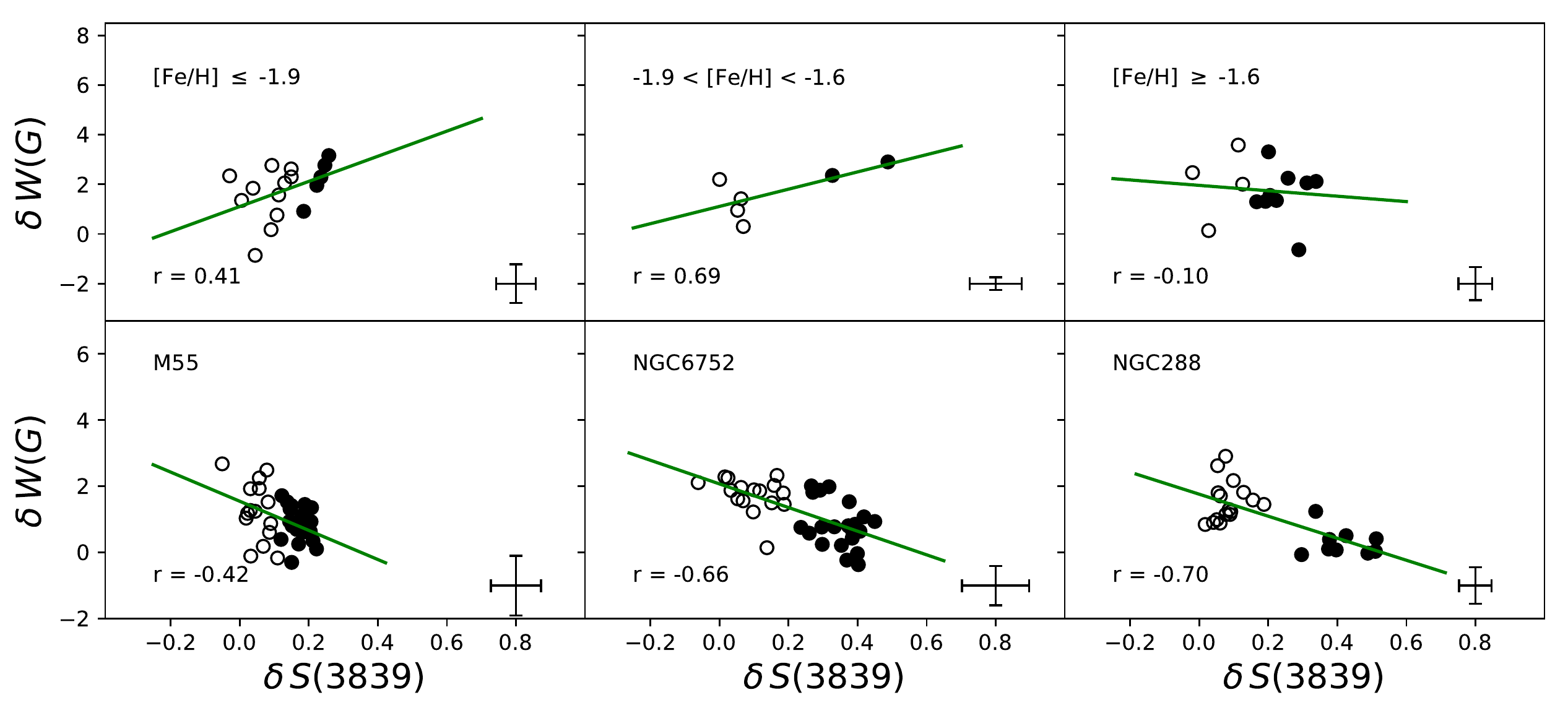}
\caption{Upper row: the dependence of $\delta S(3839)$ on $\delta W(G)$ for RGB stars of 
Sculptor with [Fe/H] $\leq$ --1.9, --1.9 $<$ [Fe/H] $<$ --1.6 and [Fe/H] $\geq$ --1.6 dex.  Lower: the same quantities for the RGB members of the GGCs M55, NGC~6752 and NGC~288. Filled and empty circles represent CN-strong and CN-weak stars as defined by the middle panels of Fig.\ \ref{wg_cn_metall} and Fig. \ref{wg_cn_clusters}, respectively. Green lines represent the best fit in each case. Correlation coefficient (r) is shown in each panel. Note the strong anti-correlation between the two parameters for GGC stars.  Error bars for each index and data set are shown in the bottom right corner in each panel.} 

\label{deltas}
\end{center}
\end{figure*}

In Fig. \ref{deltas} we show plots of $\delta W(G)$ against $\delta S(3839)$ for the three metallicity groups of Scl stars and for the three comparison GGCs.  The $\delta$ values have been measured using the appropriate lower envelope lines, and their errors are given in Table \ref{tabla_error_delta}. 
The error in each delta was calculated as a combination of the error in the index plus the uncertainty in the location of the lower envelope line. The bottom row in the Figure confirms the well-established result that in GGCs, CN- and CH-band strengths are anti-correlated.  The upper rows of the figure indicate, however, that this is evidently not the case for the Scl stars -- indeed the opposite appears to the the case in that either indices are not correlated (upper right panel), or there is a trend for the CN-strong stars to have relatively stronger CH-bands (left and middle panels)\footnote{At fixed metallicity and T$_{eff}$, the $W(G)$ and S(3839) indices, and their corresponding $\delta$ values, might naturally be expected to increase with increasing carbon abundance.  We have investigated this effect by measuring the indices on a series of synthetic spectra with varying carbon abundances.  Specifically, we adopted the parameters T$_{eff}$ = 4500 K, log $g$ = 1.0 and [Fe/H] = --1.5 dex, and calculated the indices from synthetic spectra with carbon abundance ratios [C/Fe] of --1.0, --0.5, 0.0, 0.5 and 1.0 dex.  We find that increasing the carbon abundance does increase both indices, but with a slope $\Delta W(G)$/$\Delta$S(3839) that is substantially steeper than those exhibited by the positive correlations in the upper panels of Fig.\ \ref{deltas}.  We conclude that the relations in the upper panels of Fig.\ \ref{deltas} are not driven solely by changing carbon abundances.}
This can be corroborated by the values of the correlation coefficient {\it r} which has values of 0.41, 0.69 and --0.10 for metal-poor, intermediate-metallicity and metal-rich groups, respectively.  The correlation coefficient assumes values in the range from --1 to +1, where --1 indicates  a strong negative relationship between the data and +1 indicates a strong positive relationship. 

We have sought confirmation of these results by making use of the CH($\lambda$4300) and S(3839) indices from \citet{lardo2016carbon}.  We binned their Scl stars into the same metallicity groups as used here and employed their index values to generate panels equivalent to those in the upper part of Fig.\ \ref{deltas}.  We find that the resulting $\delta CH(4300)$ versus
$\delta S(3839)$ plots bear considerable resemblance to those for our data.  In particular, as in Fig.\ \ref{deltas}, there is only a weak positive correlation between these two quantities: the correlation coefficients are $r$ = 0.38, 0.56 and 0.10 for the metal-poor, intermediate-metallicity and metal-rich groups, respectively, values that very similar to those found for our data.

\begin{table}
\caption{Errors in $\delta W(G)$, $\delta S3839$ and $\delta S4142$ for the Scl stars}
\resizebox{\linewidth}{!}
{\begin{threeparttable}
\renewcommand{\TPTminimum}{\linewidth}
\centering 
\begin{adjustbox}{max width=8.5cm}
\begin{tabular}{l c c c r} 
\hline 
\hline
\normalsize{Index} & \normalsize{[Fe/H] $\leq$ --1.9} & \normalsize{--1.9 $<$ [Fe/H] $<$ --1.6} & \normalsize{[Fe/H] $\geq$ --1.6} & \\ [0.5ex] 
\hline 

$\delta W(G)$ 		&	0.77		& 0.25		& 0.66		&\\
$\delta S3839$		&	0.05		& 0.07		& 0.04		&\\
$\delta S4142$		&	0.05		& 0.02		& 0.02		&\\

\hline 

\end{tabular}
\end{adjustbox}
\end{threeparttable}}
\label{tabla_error_delta} 
\end{table}

\begin{figure*}
\begin{center}
\includegraphics[width=16cm]{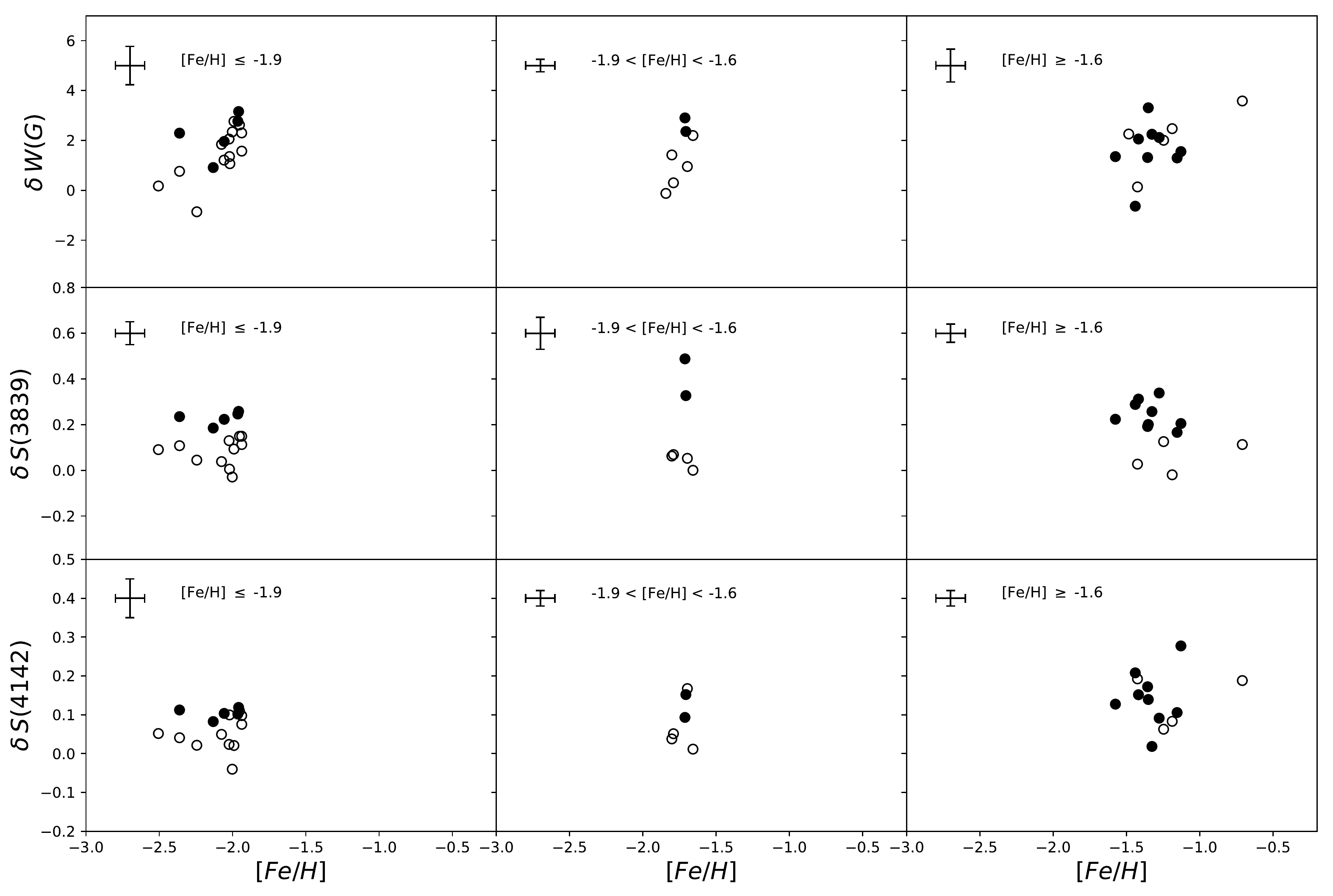}
\caption{$\delta W(G)$, $\delta S(3839)$ and $\delta S(4142)$ band strength indices versus [Fe/H] for the Scl stars.  The range of metallicities for each Scl group is indicated at the top of each panel. Filled and empty circles represent CN-strong and CN-weak stars as defined by the middle panels of Fig.\ \ref{wg_cn_metall}. Error bars are shown for the $\delta$ values (from Table \ref{tabla_error_delta}) and for the metallicity values.}
\label{comp_met}
\end{center}
\end{figure*}

In order to understand if the apparent positive correlation between CN and CH for the Scl stars could be an effect of the metallicity range in each Scl group, we have plotted the $\delta$ values for the indices against our estimation of the metallicities of the individual stars. 
We decided to use $\delta$ values in order to minimize the effects of the different temperatures and surface gravities of the Scl stars. In Fig. \ref{comp_met} no obvious trends are visible, on the contrary there are substantial ranges in all the $\delta$ values, which are similar across all three metallicity groups.  If we instead use [Fe/H] values for the stars from other sources, e.g., \citet{kirby2009multi} and \citet{starkenburg2010nir}, the outcome is essentially unaltered; we conclude that overall metallicity is not responsible for the spreads in the CN and CH-indices. 

In the next sub-sections we discuss each index in further detail and compute synthetic spectra to investigate the abundance differences implied by the range in index values.

\subsubsection{S(3839) index}

 
The $S(3839)$ index for the lowest metallicity group of Scl stars shows a range that is similar to the range shown by the M55 red giants (see Fig. \ref{wg_cn_clusters}); in both cases the range is about $\sim$0.2 mag. The average of the observational errors in this group of Scl stars is $\varepsilon \approx \pm$ 0.05 mag, which is similar in size to the GGC data for M55. Therefore a genuine dispersion in the index values of these Scl stars may be present.

For the intermediate metallicity group we have a relatively small sample of Scl stars.  However, the index values appear to have a real spread whose range is comparable to that seen for the GGC NGC~6752 -- in both cases the range in $S(3839)$ is $\sim$0.3 mag. The cluster NGC~6752 shows a very clear bimodality (see Fig.\ \ref{wg_cn_clusters}) but the limited sample size for Scl makes it difficult to establish the presence or absence of a similar bimodality.  We note also that the overlap in luminosity between the NGC~6752 and the Scl samples is limited: the Scl stars are all more luminous than $V-V_{HB}$ = --2.5 while the number of NGC~6752 stars in this range is small and no CN-strong stars were observed.
As regards the most metal-rich group of Scl stars, they appear to have a smaller range in this index than is shown by the red giants in the comparison GGC NGC~288: the range for the Scl stars is about $\sim$0.3 mag while that for the cluster is $\sim$0.45 mag.  Further, the distribution of the Scl star index values is evidently not obviously bimodal, whereas the cluster data clearly is.  This difference does not seem to be a consequence of the slightly larger errors for the Scl stars ($\varepsilon \approx \pm$ 0.05 dex)  as compared to the NGC~288 stars ($\varepsilon \approx \pm$ 0.04 dex).
Further, as for NGC~6752 and the intermediate metallicity Scl stars, there are essentially no NGC~288 stars in the same range of $V-V_{HB}$ as for the metal-rich Scl stars.  Indeed $\sim$46$\%$ of stars of the stars in the Scl metal-rich group are more metal-rich and cooler than the NGC 288 stars. 

\subsubsection{S(4142) index}
\label{S(4142)index}


The $S(4142)$ index for M55 in GGCs panels of Fig. \ref{wg_cn_clusters} does not yield any additional information about the behaviour of CN in the cluster stars, but this is not unexpected given the relatively
low metallicity of the cluster.  This is also the case for the Scl stars in the low metallicity group although there is a tendency for the stars classified as CN-strong from the $S(3839)$ index to also have higher $S(4142)$ indices.
Based on the range of $\sim$ 0.15 mag and the average index error $\varepsilon \approx~\pm$ 0.01 mag, we conclude there is a real spread in the index values for this group of Scl stars.  We note, however, that the $S(4142)$ errors for the Scl stars are notably smaller than those computed for the GGC stars despite the generally lower S/N of the Scl spectra.  This suggests that the derived S(4142) index errors for these Scl stars may be underestimated.

At intermediate metallicity, the data for the GGC NGC~6752 shows a clear separation of the CN-strong and CN-weak groups, demonstrating that at this [Fe/H] there is sufficient sensitivity to detect variations in CN-band strength through the $S(4142)$ index.  We note that the two Scl stars that were classified as CN-strong by the $S(3839)$ index values also appear relatively CN-strong in this panel.  We caution against over-interpretation of this panel, however, as there is also a Scl star with strong $S(4142)$ that is CN-weak in the $S(3839)$ panel. 
At the highest metallicities, in the GGC NGC 288 the CN-weak/CN-strong stars defined by the $S(3839)$ index are generally in their expected locations in the $S(4142)$ panel of Fig.\ \ref{wg_cn_clusters}, but the separation is not as clear cut. 
This is also the case for the metal-rich Scl group -- the CN-strong and CN-weak stars classified by the  $S(3839)$ index are intermingled in the $S(4142)$ panel.  

\subsubsection{The G band}


The first row of the Fig. \ref{wg_cn_clusters} shows in a quite convincing way the expected result for GGCs: that stars considered as CN-strong lie low in the ($W(G)$, $V-V_{HB}$) plane, consistent with lower [C/Fe] values (see Section \ref{spec_synth}).
As regards the Scl stars, in general there is little evidence for a CN/CH anti-correlation in the upper panels of Fig \ref{wg_cn_metall}, and thus there is a clear difference with respect to the GGCs.  Nevertheless, given the error estimates for the $W(G)$ values, it does appear that there is a real spread in the $W(G)$ values in all three Scl metallicity groups.  Specifically, the observed standard deviations of the $W(G)$ values are $\sigma$ = 0.96, 0.98 and 1.02 \AA, for the metal-poor, intermediate and metal-rich groups, respectively. Subtracting in quadrature the mean error in $W(G)$ ($\varepsilon$ = 0.12, 0.21 and 0.17 \AA), the implied intrinsic dispersions in $W(G)$ are then $\sigma_{int}$ = 0.95, 0.95 and 1.00 \AA\/ for the three Scl metallicity groups.
The intrinsic dispersion in the $W(G)$ values are presumably driven by intrinsic variations in [C/Fe] on the Scl red giant branch as discussed by \citet{lardo2016carbon}.

\begin{figure}
\begin{center}
\includegraphics[width=8.5cm]{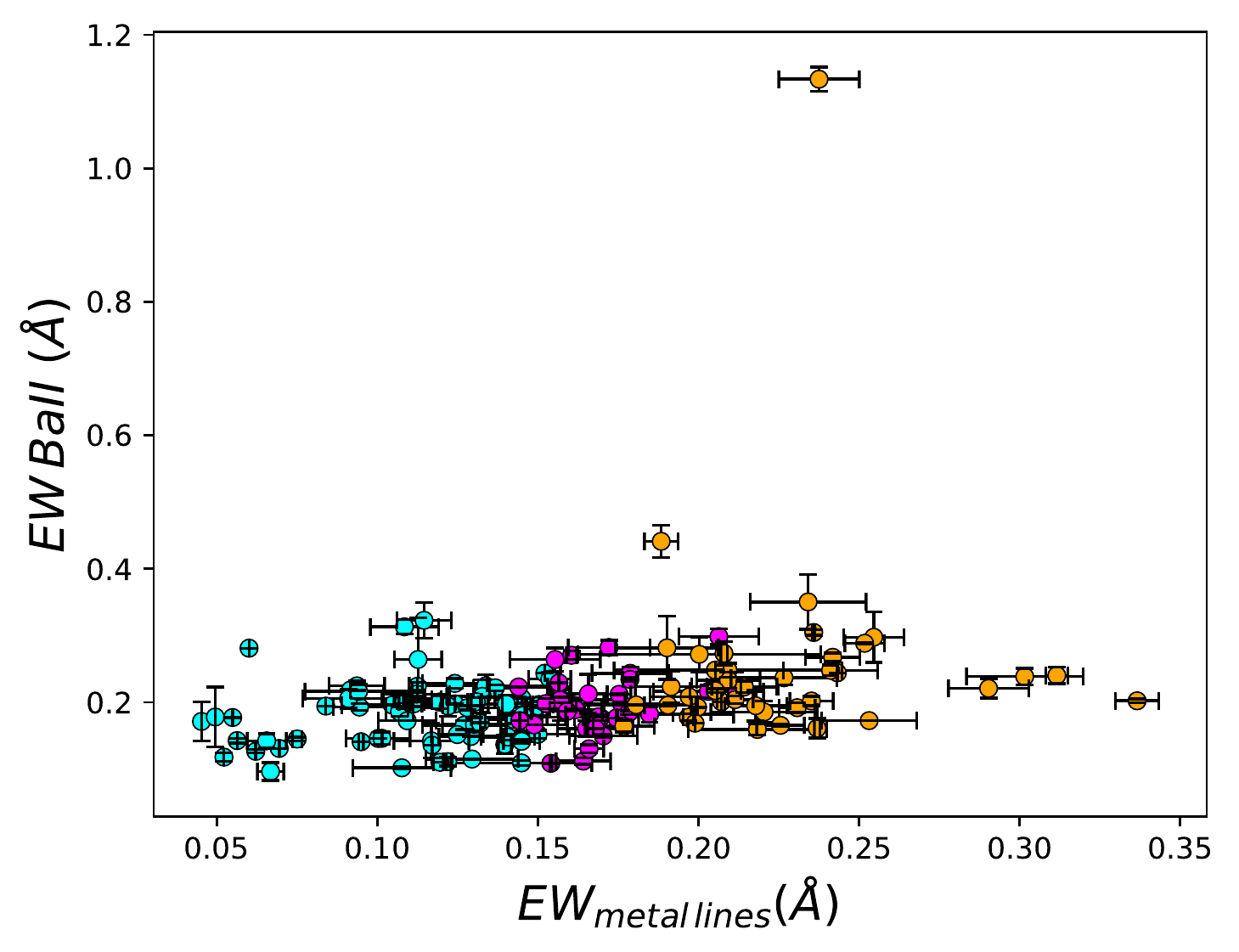}
\caption{Equivalent width of Barium {\sc ii} line at 6142 \AA\/ against equivalent width  of metal lines (12 strong lines of
calcium, iron and nickel used in the metallicity determination (See section \ref{metallicities_procedure})). For the observed stars the colour 
coding refers to different metallicity groups: cyan, magenta and orange are used for stars with [Fe/H] $<$ --1.9, --1.9 $\leq$ [Fe/H] $\leq$ --1.6 and [Fe/H] $>$ --1.6 respectively. Errors represent the  standard deviation of two measurements scaled by $\sqrt{2}$.}
\label{barium}
\end{center}
\end{figure}

Some stars in this figure seem to be both CN-strong and CH-strong, which might be an indicator of a possible CEMP classification: certainly the CEMP-s Scl star (Scl-1013644) discussed in \citet{salgado2016scl} is both CN- and CH-strong.  To investigate whether there are any other candidate CEMP-s stars in our Scl sample, we measured the strength of the $\lambda$6142\AA\/ Ba {\sc ii} line, W(Ba~{\sc ii}), in the red spectra for the full sample of Scl giants.  The results are shown in Fig.\ \ref{barium}, where it is evident that aside from the star Scl-1003644, which has a very strong Ba~{\sc ii} line \citep{salgado2016scl}, no other Scl giants in our sample possess anomalously strong Ba~{\sc ii} lines that would suggest a CEMP-s classification. Fig.\ \ref{barium} also shows that there is a correlation between the W(Ba~{\sc ii}) values and equivalent width of the combined metal lines used to determine the metallicities of the Scl stars, as expected since in general we expect the barium abundance to scale with metallicity.



\subsubsection{Spectrum synthesis}
\label{spec_synth}

A number of studies have shown that the observed CN/CH band-strength anti-correlation in GGC red giants is driven by anti-correlated changes in the [N/Fe] and [C/Fe] abundances.  Generally relative to the CN-weak stars, the CN-strong stars have increased [N/Fe] by factors of 4--10$\times$ and depletions in [C/Fe] of 2--3$\times$ for a given assumption regarding [O/Fe] \citep[e.g.,][]{schiavon2016apogee, smith2005comparison}.  To investigate the abundance ranges implied by our data, we have computed a number of synthetic spectra for pairs of [N/Fe], [C/Fe] and [O/Fe] and for $T_{eff}$, $log$\,$g$ and [Fe/H] values that follow the RGBs of the three comparison GGCs.  
The analysis was performed using the local thermodynamic equilibrium (LTE) spectrum synthesis program MOOG \citep{sneden1973nitrogen, sobeck2011abundances} and ATLAS9 model atmospheres \citep{castelli2003new}.  The solar values used were $T_{eff, \odot}$ = 5790 K,  $log\, g _{\odot}$ = 4.44,  and $M_{bol, \odot}$ = 4.72.  The effective temperatures $T_{eff}$ were obtained using the reddening corrected RGB colours and the $T_{eff}$-colour calibration of \citet{alonso1999effective}.   The bolometric magnitudes were derived from the RGB $V$ magnitudes and the reddenings and distance moduli listed in the most recent version of the \citet{harris1996catalog} catalogue for the GGCs (see Table \ref{tabla_harris}).  For Scl a reddening of $E(B-V)$ = 0.018 and a distance modulus of 19.67 \citep{pietrzynski2008araucaria} was used.  We assumed $M_{*}$ = 0.8 solar masses for the RGB stars and then calculated the surface gravities from the stellar parameters.  Finally, the synthetic spectra were smoothed to the resolution appropriate for each set of observations.

\begin{table}
\caption{[C/Fe], [N/Fe] and [O/Fe] abundances used for in synthetic spectra calculations}
\resizebox{\linewidth}{!}
{\begin{threeparttable}
\renewcommand{\TPTminimum}{\linewidth}
\centering 
\begin{tabular}{l c c c r} 
\hline 
\hline
\normalsize{} & \normalsize{M55} & \normalsize{NGC 6752} & \normalsize{NGC 288} &\\ [0.5ex] 
\hline 
$[C/Fe]{\tiny _{CN-weak\star}}$ &	--0.7	 	&	--0.6		&	--0.5		&\\
$[N/Fe]{\tiny _{CN-weak\star}}$	  &	--0.5		&	--0.6		&	--0.5		&\\
$[O/Fe]{\tiny _{CN-weak\star}}$	  &	 0.4		&	 0.4		&	 0.4		&\\
 & & & &\\
$[C/Fe]{\tiny _{CN-strong\star}}$ &	--1.2		&	--1.1		&	--1.0		&\\
$[N/Fe]{\tiny _{CN-strong\star}}$ &	 0.5		&	 0.4		&	 0.5		&\\
$[O/Fe]{\tiny _{CN-strong\star}}$ &	 0.0		&	 0.0		&	 0.0		&\\
\hline 

\end{tabular}
\end{threeparttable}}
\label{variations} 
\end{table}

Once the set of synthetic spectra were available we measured the $W(G)$, $S(3839)$ and $S(4142)$ indices.  The resulting band-strength index values were then used to define the shaded regions in the panels of Fig.\  \ref{wg_cn_clusters}: in the $S(3839)$ and $S(4142)$ panels the upper boundary is for the
assumed abundances for the CN-strong stars, and the lower boundary is for the CN-weak abundances (see Table \ref{variations}).  The reverse is the case for the $W(G)$ panel.  Generally the agreement between the synthetic spectra predictions and the location of the observed cluster populations is satisfactory, confirming the expected abundance differences between the populations.  We note that it was necessary to adopt [C/Fe] $\approx$ --0.6 for the CN-weak stars (see Table \ref{variations}) as the predicted $W(G)$ values were substantially too large with [C/Fe] = 0.0 dex.  This low value of [C/Fe] is not inconsistent with the predictions of evolutionary mixing on the RGB as discussed, for example, by \citet{placco2014carbon}.   

We note also that our values for the nitrogen abundances for the CN-weak GGC red giants are comparable with other determinations in the literature.  For example, \citet{yong2008nitrogen} found for NGC~6752 values of nitrogen from [N/Fe] = $-0.43$ upwards, while for NGC~7078 (M15) which has [Fe/H] = $-2.37$, \citet{trefzger1983carbon} found [N/Fe] $\approx$ $-0.6$ dex.   Similarly, \citet{suntzeff1981carbon} found [N/Fe] = $-0.4$ for CN-weak stars in NGC~5272 (M3) and NGC~6205 (M13) both of which have [Fe/H] $\approx$ $-1.5$ dex.   The specific [C/Fe], [N/Fe] and [O/Fe] abundances adopted for the CN-weak and CN-strong populations in each GGC are listed in Table \ref{variations}, noting that the [O/Fe] values used are assumptions consistent with observations of O-strong (Na-weak) and O-weak (Na-strong) stars in GGCs as we have no means to directly determine oxygen abundances from our data.

To investigate the extent to which the Scl stars show similar [C/Fe] and [N/Fe] variations to the GGC stars,
we smoothed the synthetic spectra to the lower resolution of the Scl blue spectra and recomputed the values the band-strength indices.  The outcome is shown as the shaded regions in the panels of Fig.\ \ref{wg_cn_metall}.  For the $S(3839)$ index, the majority of the Scl stars, in all the metallicity groups, lie below the lower boundary of the shaded region, with only the CN-strong objects reaching into the shaded region.  Nevertheless the range in the Scl index values is approximately comparable to the width of the shaded region.  Similarly, in the $W(G)$ panels, again the Scl stars lie below the shaded region in all three metallicity groups, although, like the situation for $S(3839)$, the observed range of the Scl data points is comparable to, or perhaps somewhat larger than, the width of the shaded region.  Taken at face value this would seem to indicate that the [C/Fe] values for these luminous Scl stars, which lie near the RGB-tip, are yet lower than the [C/Fe] $\approx$ --1.0 dex assumed for the GGC CN-strong stars.  Whether this additional carbon depletion is sufficient to reduce the values of the synthetic $S(3839)$ indices into the observed region is unclear as CN-band strength is sensitive to both [C/Fe] and [N/Fe].  The panels for the $S(4142)$ index complicate things further for here the Scl stars show the opposite of the situation for $S(3839)$ -- in all metallicity bands the Scl stars lie above the shaded region, and the observed range is larger than the width of the shaded regions.

\subsection{C and N abundances via spectrum synthesis of representative Scl stars}
\label{synthCN}

In order to quantify and confirm the interpretation of Figs \ref{wg_cn_metall} and \ref{deltas} we have carried spectrum synthesis calculations to determine C and N abundance estimates for a representative subset of the Scl stars in each metallicity group.  The 16 stars chosen, which are listed in Table \ref{Tabla_parameters2}, encompass the full range of S(3839), S(4142) and $W(G)$ values, as well as [Fe/H], in each metallicity group.  In order to perform the synthetic spectrum calculations, however, we need to assume [O/Fe] values for the stars.  As indicated in Table \ref{Tabla_parameters2}, we have chosen to assume [O/Fe] = +0.4 for the CN-weak stars and [O/Fe] = 0.0 for the CN-strong stars, the former value being consistent with GGC first generation and halo field stars and the latter value being consistent with the [O/Fe] values for second generation stars in GGCs.  However, this assumption does not strongly influence the outcome: if instead we employ [O/Fe] = +0.4 for the CN-strong stars, the derived [C/Fe] values are approximately 0.1 dex smaller than for the [O/Fe] = 0.0 case, while the derived [N/Fe] values are $\sim$0.25 dex larger with [O/Fe] = +0.4 compared to the [O/Fe] = 0 case.

\begin{figure}
\begin{center}
\includegraphics[width=8.5cm]{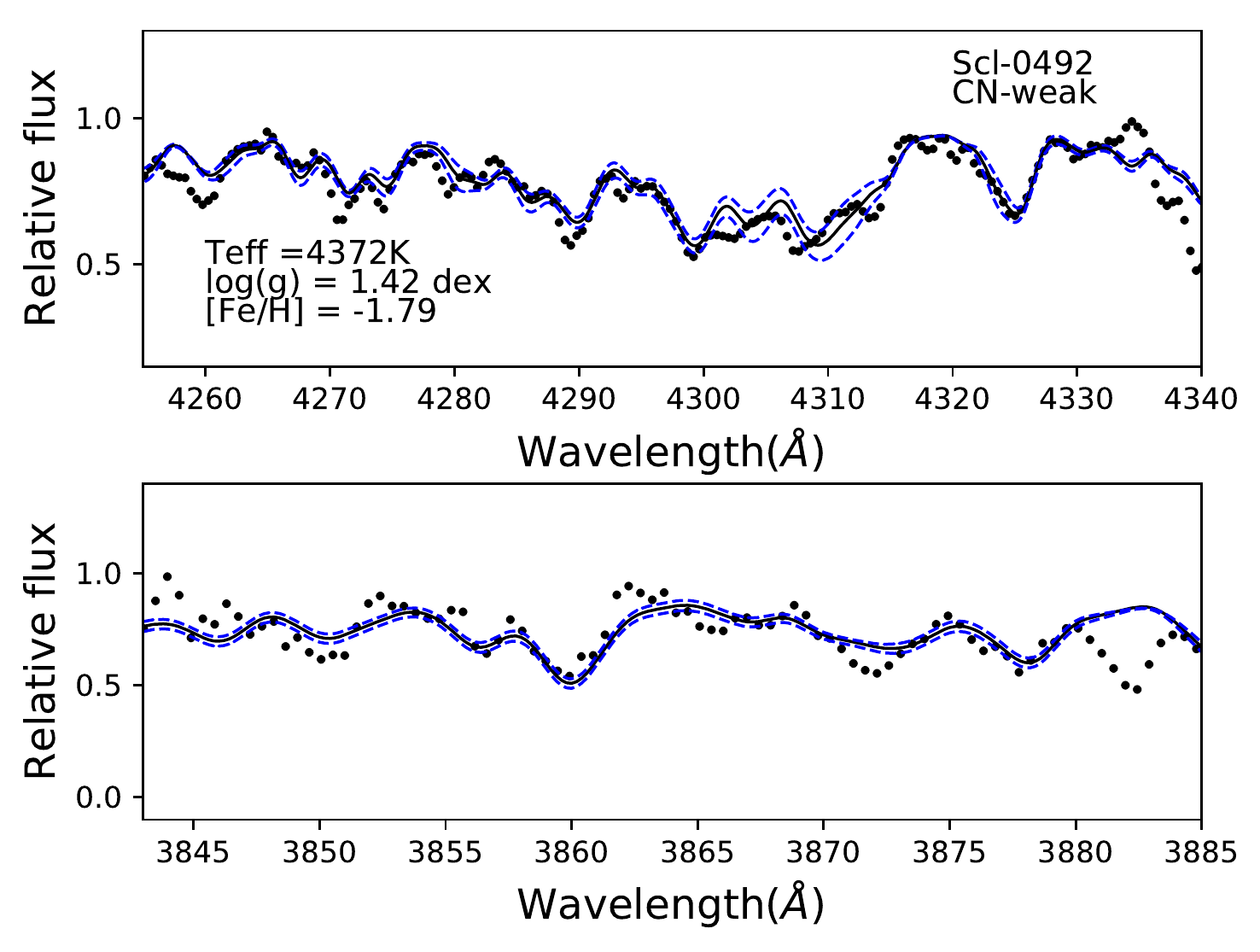}
\caption{Spectrum synthesis of CH (top panel) and CN (bottom panel) features for 
Scl-0492 (CN-weak). In both panels the observed spectrum is represented by solid dots. Upper panel: the solid line represents the best abundance fit which has [C/Fe] = --0.87 assuming [O/Fe] = +0.4. The dashed blue lines show [C/Fe] values $\pm$0.16 dex about the central value. Lower panel: the solid line represents the best abundance fit which has [N/Fe] = --0.6 and [C/Fe] = --0.87, assuming [O/Fe] = +0.4. The dashed blue lines show nitrogen values $\pm$0.3 dex from the best fit. } 
\label{synth_comp} 
\end{center}
\end{figure}

\begin{figure}
\begin{center}
\includegraphics[width=8.5cm]{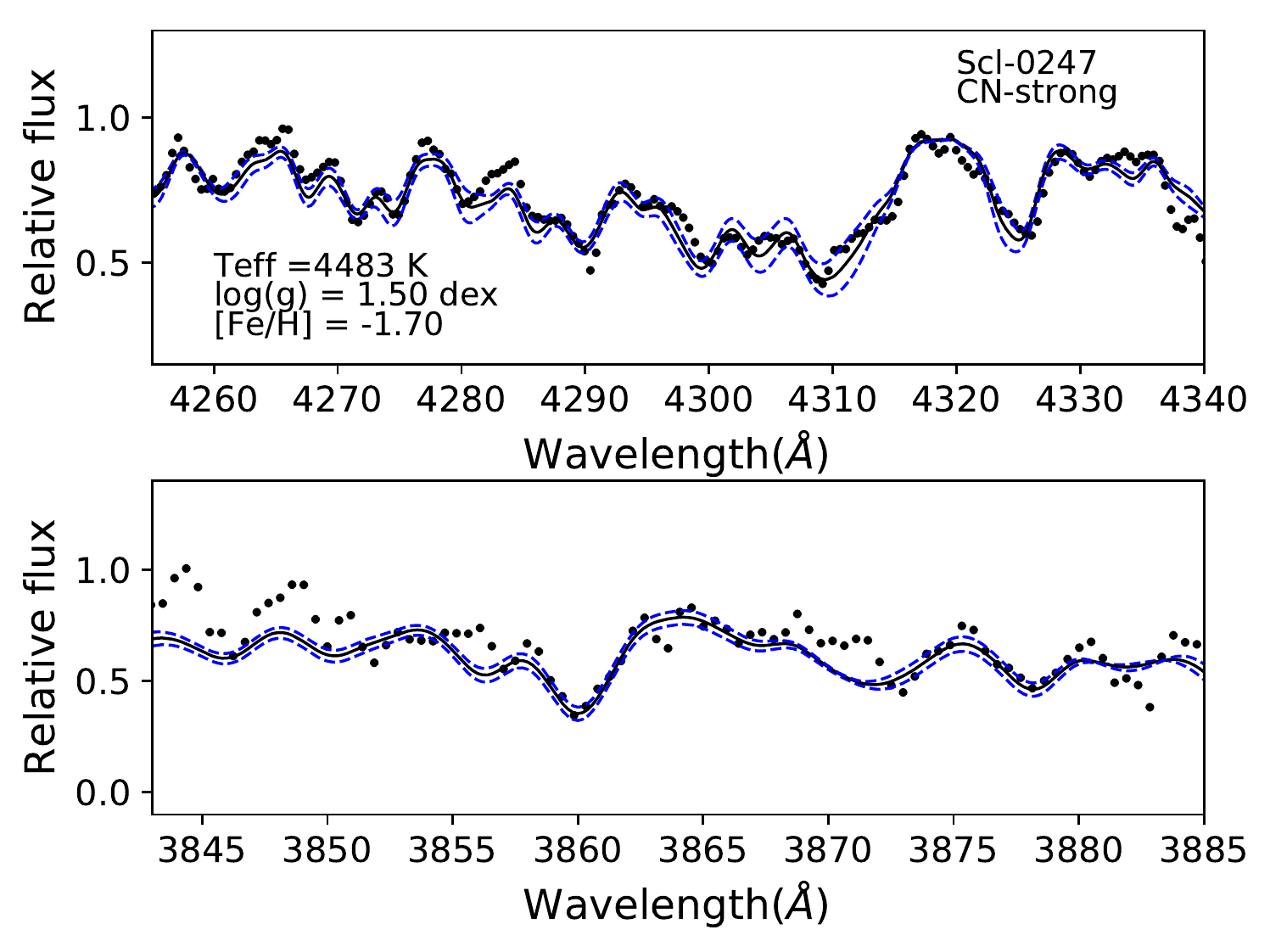}
\caption{Spectrum synthesis of CH (top panel) and CN (bottom panel) features for 
Scl-0247 (CN-strong). In both panels the observed spectrum is represented by solid dots. Upper panel: the solid line represents the best abundance fit  which has [C/Fe] = --0.83 assuming [O/Fe] = 0.0. The dashed blue lines show carbon values $\pm$0.17 dex about the central value. Lower panel: the solid line represents the best abundance fit with  [N/Fe] = +0.00 and [C/Fe] = --0.83, assuming [O/Fe] = 0.0. The dashed blue lines show nitrogen values $\pm$0.3 dex from the best fit.} 
\label{synth_comp247} 
\end{center}
\end{figure}

The synthetic spectrum calculations were carried as described in subsection \ref{spec_synth} and the stellar parameters adopted for each modeled star are given in Table \ref{Tabla_parameters2}.  The procedure followed was to first determine the carbon abundance (assuming [O/Fe] = 0.0 or [O/Fe] = +0.4) by minimizing the residuals between the observed and synthetic spectra in the region of the G-band of CH ($\lambda\, \approx$ 4300 \AA).  The nitrogen abundance was the determined by considering the comparison of the synthetic and observed spectra in the wavelength interval 3840 -- 3885 \AA\/ using the value of [C/Fe] determined from the G-band and the assumed oxygen abundance.  The process is illustrated in Figs \ref{synth_comp} and \ref{synth_comp247} for the intermediate-metallicity group stars Scl-0492 (CN-weak) and Scl-0247 (CN-strong).  The outcome is a difference in [N/Fe] between the two stars of a factor of approximately 4 (CN-strong star has higher [N/Fe]), but the difference in the derived [C/Fe] is essentially zero given the $\pm$0.15 dex uncertainties in the [C/Fe] determination.   The derived [C/Fe] and [N/Fe] abundances for the 16 stars analysed are given in Table \ref{Tabla_parameters2}.  

The total errors,  $\sigma_{total}$, in the derived [C/Fe] and [N/Fe] values were calculated as a combination of the errors from uncertainties in the stellar parameters ($\sigma_{SP}$) with additional sources of error ($\sigma_{fit}$). The quantity $\sigma_{SP}$ was estimated by repeating the abundance analysis for the star Scl-1490 (assumed to be representative of the sample --- see Table \ref{Tabla_parameters2}) varying the atmospheric parameters by $T_{eff} = \, \pm 100$K, $log \, g = \pm 0.2$, $[M/H] = \pm 0.4$ and $\xi = \pm 0.2$ km~s$^{-1}$. The quantity $\sigma_{fit}$ comes from the uncertainty in the derived abundances resulting from the comparison of the observed spectrum with model spectra at different abundances.   As the derived carbon abundance is dependent on the assumed oxygen abundance, for $\sigma_{fit\,C}$ we have added (in quadrature) a further $\sigma_{fit}= \pm 0.1$ to allow for the uncertainty in adopted oxygen abundances.   Similarly, since the determination of the N abundance from the CN features depends of the derived carbon abundance, we have added (in quadrature) to $\sigma_{total}$ for nitrogen the value of $\sigma_{total}$ for carbon.  Typical values of $\sigma_{total}$ are 0.16 dex for [C/Fe] and 0.29 dex for [N/Fe].  

\begin{figure*}
\begin{center}
\includegraphics[width=16cm]{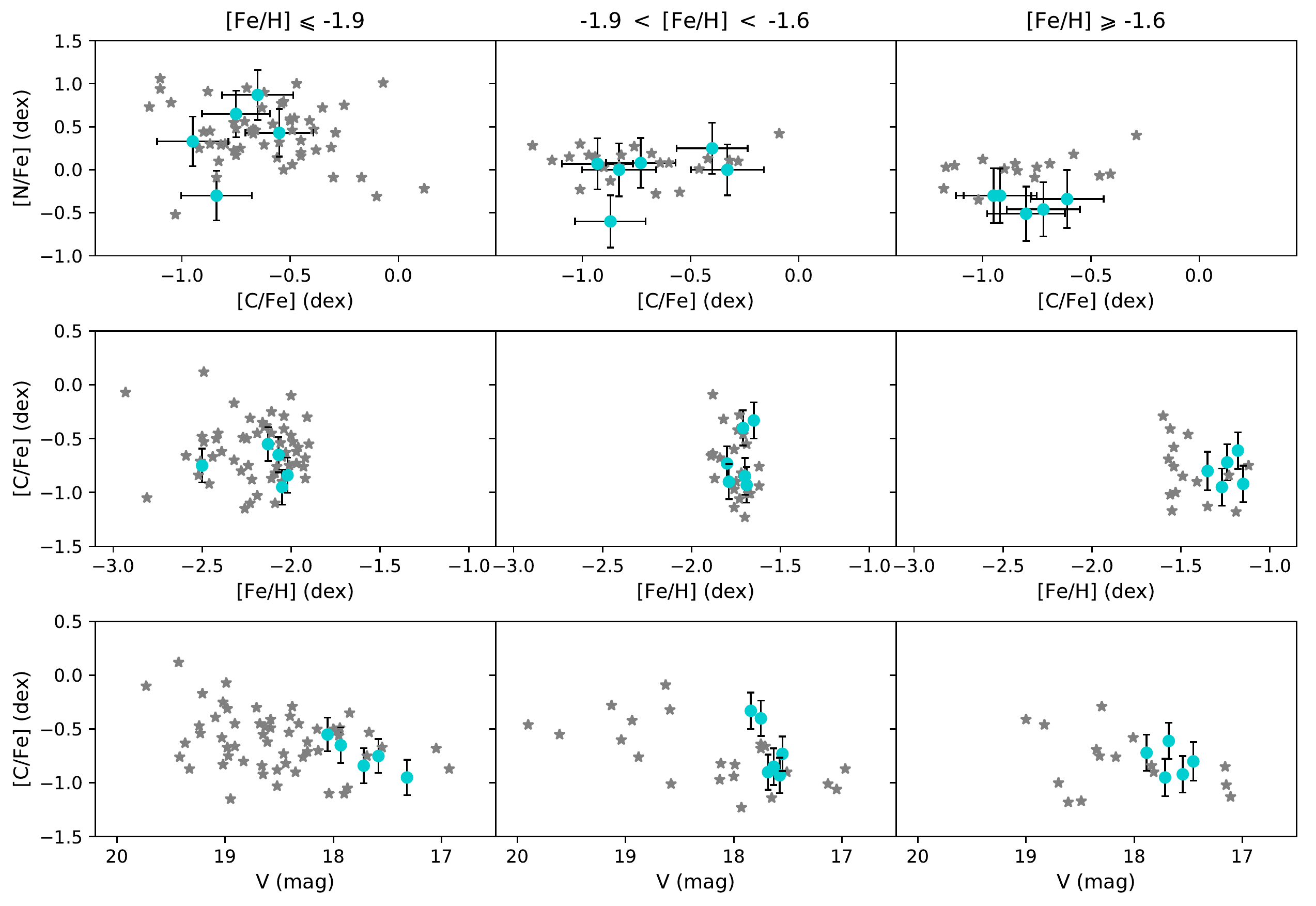}
\caption{Comparison between the Sculptor stars from \citet{lardo2016carbon} (grey-stars) and a representative sub-set of our sample (cyan filled circles with error bars). The stars have been separated into three group of metallicities: [Fe/H] $\leq$ --1.9, --1.9 $<$ [Fe/H] $<$ --1.6 and [Fe/H] $\geq$ --1.6 dex. Upper row: the relation between [C/Fe] and [N/Fe] abundances. Middle row: [C/Fe] abundances as function of metallicities [Fe/H]. Lower row: [C/Fe] against $V$ magnitude. 
} 
\label{synth_CN}
\end{center}
\end{figure*}

Our results for [C/Fe] and [N/Fe] are shown in Fig. \ref{synth_CN} where we have also plotted the C and N abundance ratios derived by 
\citet{lardo2016carbon} for their sample of Scl stars.  Unfortunately, we cannot directly compare on a star-by-star basis our results with those of \citet{lardo2016carbon} because the majority of their observed stars are fainter than ours: there is a small overlap in 
$V$ but there are no stars in common.
The two sets of data are in excellent agreement.  The upper row shows that for the intermediate and metal-rich groups there is no evidence for a either positive or negative correlation between 
[C/Fe] and [N/Fe], while an anti-correlation between [N/Fe] and [C/Fe] may be present in the panel for the most metal-poor group, though there is considerable variation in both abundance ratios.  This contrasts with the situation for GGCs where a strong anti-correlation between [C/Fe] and [N/Fe] is generally present.  The middle row shows that there is a large dispersion in [C/Fe] at all metallicities while the bottom row shows that the largest carbon depletions are generally found in the stars at the highest luminosites.
Overall, our synthesis results are consistent with those of \citet{lardo2016carbon}, particularly that there is an intrinsic range in carbon abundance which is present at all metallicities and which cannot be entirely explained by evolutionary mixing effects.

\begin{table}
\caption{Stellar parameters and derived C, N abundances for a representative sub-sample of the data}
\resizebox{\linewidth}{!}
{\begin{threeparttable}
\renewcommand{\TPTminimum}{\linewidth}
\centering 
\begin{tabular}{l c c c c c c c r} 
\hline 
\hline
\normalsize{ID}  & \normalsize{T$_{eff}$} & \normalsize{log(g)} &  \normalsize{$\xi_{t}$} & \normalsize{[Fe/H]} &  \normalsize{[C/Fe]}&  \normalsize{[N/Fe]}& \normalsize{[O/Fe]}&\\ [0.5ex] 
\normalsize{} &  \normalsize{K} & \normalsize{dex} & \normalsize{km~s$^{-1}$} & \normalsize{dex}  & \normalsize{dex} & \normalsize{dex} & \normalsize{dex} &\\ [0.5ex] 
\hline 

Scl-0272	&	         4558	&	1.48	&	1.74	&	--2.50	&	--0.75	&	0.65	&	0.4		\\	
Scl59		&	4674	&	1.24	&	1.82	&	--2.13	&	--0.55	&	0.43	&	0.0		\\	
Scl81		&	4635	&	1.30	&	1.79	&	--2.07	&	--0.65	&	0.87	&	0.4		\\        
Scl38		&	4422	&	1.65	&	1.68	&	--2.05	&	--0.95	&	0.33	&	0.0		\\        
Scl-1490	&	         4582	&	1.41	&	1.76	&	--2.02	&	--0.84	&    --0.30	&	0.4		\\	
 & & & & & & & \\
11\_1\_6268	&	4493	&	1.52	&	1.72	&	--1.80	&	--0.73	&	0.08	&	0.4		\\	
Scl-1020	&	         4518	&	1.43	&	1.75	&	--1.71	&	--0.40	&	0.25	&	0.0		\\       
Scl-0492	&	         4372 &	1.42 &	TBD &       --1.79        &       --0.87        &    --0.6   &       0.4              \\	
Scl-0247	& 	         4483	 &	1.50	&	TBD &       --1.70        &      --0.83         &      0.0    &       0.0             \\       
Scl-0268	&	         4518	&	1.50	&	1.73	&	--1.69	&	--0.93	&	0.07	&	0.4		\\	
Scl23		&	4814	&	1.26	&	1.81	&	--1.65	&	--0.33	&	0.00	&	0.4		\\	
 & & & & & & & \\	
10\_8\_4149	&	4327	&	1.65	&	1.68	&	--1.35	&	--0.8	         &    --0.51	&	0.0		\\	
Scl20		&	4406	&	1.50	&	1.73	&	--1.27	&	--0.95	&    --0.30	&	0.0		\\       
Scl-0437	&	         4476	&	1.40	&	1.76	&	--1.24	&	--0.72	&    --0.46	&	0.4		\\	
Scl-0276	&	         4687	&	1.38	&	1.77	&	--1.18	&	--0.61	&    --0.34	&	0.4		\\	
Scl-0233	&	         4396	&	1.58	&	1.71	&	--1.15	&	--0.92	&    --0.30	&	0.0		\\	
	
\hline 
\end{tabular}
\label{Tabla_parameters2} 
\end{threeparttable}}
\end{table}

\subsection{Sodium}
\label{na_sect}

Determinations of the sodium abundance in GGCs, and its star-to-star variation, stem principally from the analysis of high dispersion spectra, which allow the measurement of relatively weak Na lines.  Such high dispersion spectra generally also allow the measurement of abundances for oxygen, and thus investigation of the extent of the Na/O anti-correlation \citep[e.g.,][]{carretta2009anticorrelation, carretta2016spectroscopic}.
Indeed the Na/O anti-correlation is one of the key features of the GGC light element abundance anomaly phenomenon.  Here we adopt a different approach in that we use the strengths of the Na~D lines in intermediate resolution spectra as a measure of sodium abundance, and then seek evidence for a Na/N correlation using the CN-band strengths as an indicator of the N abundance.  Since in GGCs the stars that are depleted in oxygen are enhanced in N, investigating a Na/CN correlation is equivalent to exploring the Na/O anti-correlation.  This is not a new approach -- earlier studies have shown that the strengths of the Na~D lines are stronger in CN-strong GGC red giants compared to CN-weak 
stars \citep{cottrell1981, Dacosta_coll, norris1985sodium}.  

As regards Scl, the investigation of the existence of any relation between CN-band and Na~D line strengths is vital: CN-band variations on their own are not sufficient evidence for the presence of GGC-like abundance anomalies as they may result from variable amounts of evolutionary mixing on the giant branch.  A direct connection between the CN-band and Na~D line strengths, on the other hand, would be strong evidence in favour of the occurrence of the GGC abundance anomaly phenomenon in this dSph.


As a first step, we have determined the equivalent widths, denoted by $W(NaD)$, of the Na~D lines in the spectra of the RGB stars in our set of comparison GGC, using the approach outlined in Section \ref{strengths}.  For both NGC~288 and NGC~6752 the cluster radial velocity is not sufficient to allow a clear separation of the stellar Na~D lines from any interstellar component, assumed to have a velocity near zero, at the resolution of the red spectra.  However, the reddenings for these clusters are low \citep{harris1996catalog} (see Table \ref{tabla_harris}) so that any interstellar contribution to the measured Na~D line strengths in the cluster red giant spectra can be assumed to be negligible.  This is not the case for the cluster M55 which is both more metal-poor and more highly reddened than NGC~288 and NGC~6752.  Fortunately, the radial velocity of this cluster is sufficiently high that the stellar and interstellar Na~D lines are clearly separated in the M55 red giant spectra, allowing unambiguous measurement of the stellar lines. 

\begin{figure}
\begin{center}
\includegraphics[width=8.5cm]{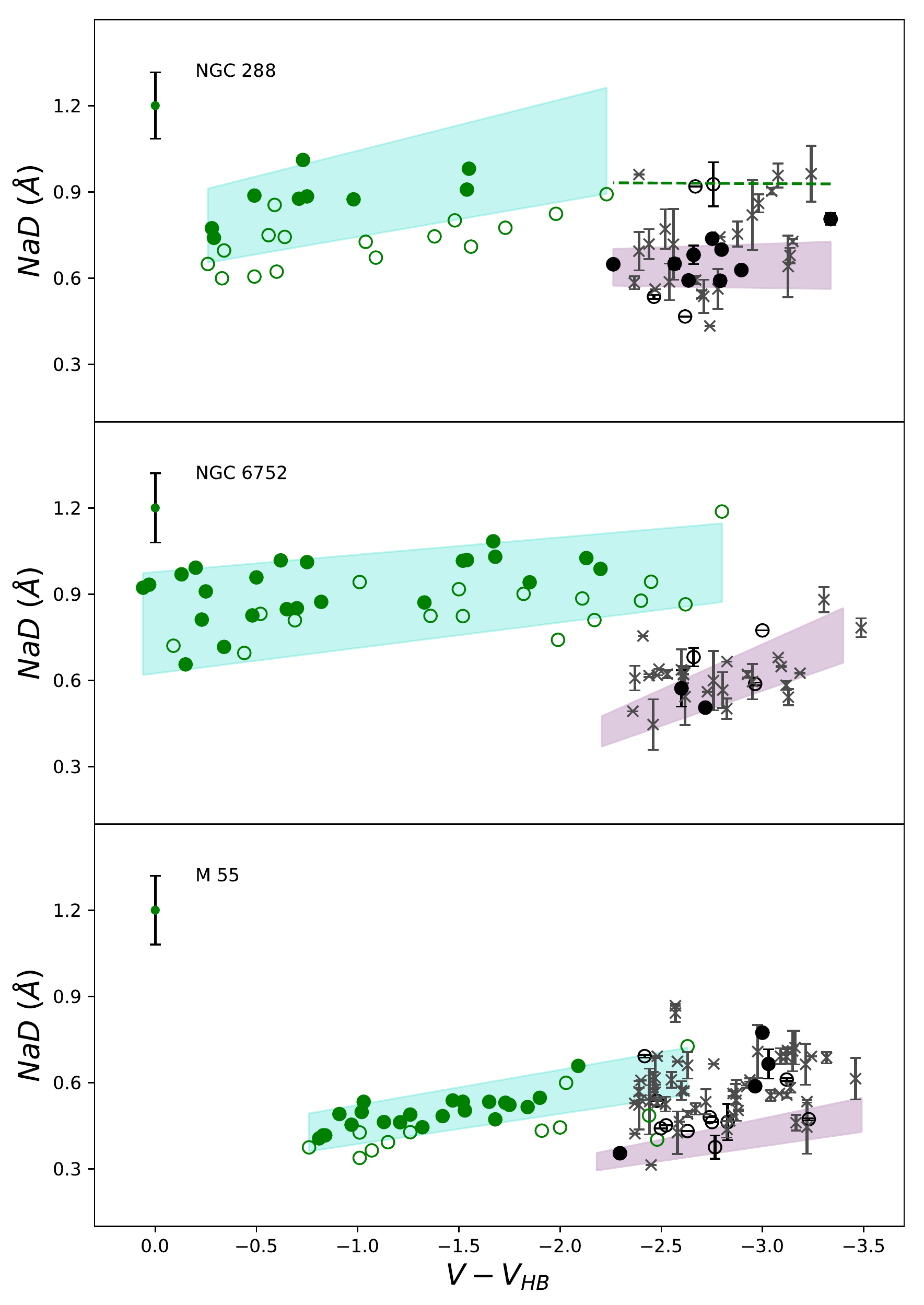}
\caption{The relation between CN-band and Na~D line strengths for RGB stars in the comparison GGCs and in Scl.  The three panels are for the three GGCs and for Scl stars in the corresponding metallicity groups, with the highest metallicity in the top panel and the lowest in the bottom panel.  The GGC stars are shown as green circles, filled for CN-strong stars and open for CN-weak stars.  A typical error bar for the GGC $W(NaD)$ measurements is shown upper left of each panel.  The green shaded region represents the range in $W(NaD)$ for a range in [Na/Fe] of 0.5 dex derived from spectrum synthesis calculations.  The Sculptor stars with blue spectra are shown as black circles, filled for CN-strong and open for CN-weak.  Grey x-symbols are used for the Scl stars that lack blue spectra. Individual uncertainties
are shown for each star.  The pink shaded region also shows the extent of a 0.5 range in [Na/Fe], but with a lower [Na/Fe] for the upper and lower limits than for the GGCs.  See text for details.  The green dashed line in the upper panel represents [Na/Fe] = 0.0 at [Fe/H] = --1.25 dex.}

\label{na}
\end{center}
\end{figure}

The results are shown in Fig.\ \ref{na} where, particularly for the more metal-rich clusters, it is apparent that the CN-strong stars, as anticipated, generally have larger $W(NaD)$ values than the CN-weak stars of similar luminosity.  The typical error in the W(NaD) values was calculated in the same way as for the indices discussed in section \ref{results cn ch na}.  Specifically, for GGCs stars, we use the rms from a fit to the CN-weak population in the ($W(NaD), V-V_{HB}$) plane.  This yields typical errors for the GGC $W(NaD)$ values of 0.11 \AA, 0.12 \AA\/ and 0.12 \AA\/ for the clusters NGC~288, NGC~6752 and M55, respectively. We have then used synthetic spectra calculations to estimate the extent of the variations in [Na/Fe] present in our GGC spectra.  The procedure was similar to that used in the section \ref{spec_synth}:  we calculated synthetic spectra for a number of ($T_{eff}, log g$) pairs appropriate to the cluster RGBs and with different values of [Na/Fe].  Based on the results of \citet{carretta2009anticorrelation} and \citet{carretta2016spectroscopic} for example, we assumed a range of $\sim$0.5 dex in [Na/Fe] as encompassing the extent of the [Na/Fe] variations present in the GGCs.  The synthetic spectra were smoothed to the observed resolution and the corresponding Na~D strengths determined.  The results are shown as the shaded bands in Fig.\ \ref{na}, where the upper boundary represents the higher value of [Na/Fe] adopted and the lower boundary the reverse.  The actual values of the upper and lower [Na/Fe] limits employed for each cluster are given in the upper section of Table \ref{tabla_na_ab}.   These values were determined by ensuring that the corresponding shaded regions in Fig.\ \ref{na} matched the location of the GGC stars.  For example, the limits on [Na/Fe] employed for NGC~288 and NGC~6752 are --0.1 and 0.4, and 0.1 and 0.6 dex, respectively, values that are entirely consistent with those based on high dispersion spectroscopy \citep{carretta2009anticorrelation} where the full range for the observed [Na/Fe] values in the NGC~288 sample is 0.05 $\leq$ [Na/Fe] $\leq$ $\sim$0.7 dex and is
0.1 $\leq$ [Na/Fe] $\leq$ $\sim$0.75 dex for NGC~6752 sample. In the case of M~55  \citet{carretta2009anticorrelation} give a range  0.1$\leq$ [Na/Fe] $\leq$ $\sim$0.7 dex  while our estimation reveals a similar range but at a lower overall [Na/Fe]: --0.5 to 0.0 dex.


For the Scl stars we followed a similar procedure, retaining the same metallicity groups as first discussed in Section \ref{metallicities_procedure}.  In particular, we note that the radial velocity of Scl is sufficiently large that any interstellar component would be distinguished from the stellar lines, but in fact we see little evidence for any interstellar component in the Scl spectra consistent with the low foreground reddening.  The measured $W(NaD)$ for the Scl giants for which we have blue spectra are shown in the panels of Fig.\ \ref{na}, with the CN-strong stars (see middle row panels of Fig.\ \ref{wg_cn_metall}) plotted as black filled circles and CN-weak stars as black open circles.  Red spectra are available for a substantially larger number of Scl members than those with blue spectra, and we show the $W(NaD)$ values for these additional stars as grey x-signs in Fig.\ \ref{na}.  The typical error for the Scl $W(NaD)$ values is given, as before, by the standard deviation of the differences between measurements on the two sets of spectra, scaled by $\sqrt{2}$.  Considering first the stars for which a CN-classification is possible, it is clear that, unlike the situation for the GGC stars, there is no indication that the CN-strong Scl stars have preferentially larger $W(NaD)$ values than their CN-weak counterparts.  The possible exception is in the most-metal poor panel where the 2 most luminous CN-strong stars have somewhat larger $W(NaD)$ values than the CN-weak stars at similar $V-V_{HB}$, although these stars are not obviously distinguished when considering the full Scl sample in this metallicity group.  We caution, however, that at the lowest metallicities the relative uncertainty in the $W(NaD)$ measurements is comparable to the size of the apparent excess in $W(NaD)$.  Overall, our results strongly imply that the GGC light element abundance anomaly is not prevalent in Scl, a result that is consistent with those from high dispersion spectroscopy that looked explicitly for the O/Na anti-correlation in Scl, and which did not find any support for its existence \citep[e.g.][]{geisler2005sculptor}.  \citet{norris2017populations} reached the same conclusion for the Carina dSph.

Two other effects are also immediately evident from Fig.\ \ref{na}.  First, whether considering just the Scl stars with blue spectra or the entire sample, the median $W(NaD)$ for the Scl stars is significantly lower than for the GGC comparison stars, even allowing for the incomplete overlap in luminosity.  The offset is particularly marked for the intermediate-metallicity stars in the middle panel, but it is also present in the upper panel.  It is less marked in the lowest abundance panel but again the $W(NaD)$ values are least certain for this Scl group.  Second, considering the full sample of Scl stars, there are small number that have $W(NaD)$ values significantly larger than the bulk of the population, and such stars appear present in all three metallicity groups.  This would suggest that there are some Scl stars that have larger [Na/Fe] values than the majority of stars with similar metallicities.

To investigate these effects we have again used synthetic spectrum calculations as for the GGC stars.  The pink shaded regions in Fig.\ \ref{na} show the range in $W(NaD)$ values for a range in [Na/Fe] of 0.5 dex, as for the GGCs.  However, in the Scl case, at least for the two more metal-rich groups, the [Na/Fe] limits are required to be at least $\sim$0.6 dex lower than the corresponding values for the GGC stars in order to reproduce the Scl line strengths.  This is less obviously the case for the most metal-poor group of Scl stars.  Here, the majority of Scl stars again have weaker Na~D line strengths than would be inferred from extrapolating the GGC M55 data, but the synthetic spectrum predictions for the same [Na/Fe] range as for the two more metal-rich groups do not encompass the Scl $W(NaD)$ values.  Whether this is a result of the larger relative uncertainty in the Scl $W(NaD)$ values at low [Fe/H], or an unknown systematic effect is unclear but, interpreted literally, the data would suggest both that the range in [Na/Fe] for these Scl stars is larger than the 0.5 dex used in the synthetic spectra calculations, and that the typical value of [Na/Fe] is higher than for the two more metal-rich groups.

Overall, the relative weakness of the Scl Na~D line strengths compared to the GGC stars at all metallicities is an outcome consistent with the high dispersion spectroscopic results of \citet{shetrone2003vlt} and \citet{geisler2005sculptor}.  They find sub-solar values of [Na/Fe] for Scl red giants: typically [Na/Fe] $\approx$ --0.55, in marked contrast to (first generation) GGC stars which have [Na/Fe] $\approx$ 0.1 dex. 
\citet{shetrone2003vlt} have found similar depletions relative to solar values for [Na/Fe] in other dSphs (Carina, Fornax, Leo~I), while the extensive study of Carina red giants of \citet{norris2017populations} finds $\langle$[Na/Fe]$\rangle$ = --0.28 $\pm$ 0.04 dex and notes that this is significantly below the typical [Na/Fe] $\approx$ 0.0 in Galactic halo (and GGC first generation) stars.  \citet{letarte2010} and \citet{lemasle2014} have shown that this is also the case in the Fornax dSph.



We now consider the panels of Fig.\ \ref{na} in more detail as regards the Scl stars.  In the top panel there is no clear separation between the Scl CN-strong and CN-weak stars and the majority of the Scl points in the full sample fall within or close to the pink band.  Indeed for the majority of the Scl stars the dispersion in the $W(NaD)$ values is consistent with a single [Na/Fe] value of order [Na/Fe] = --0.7.  There are, however, a small number of Scl stars that lie at $W(NaD)$ values substantially above the majority of the data points.  There are two possible reasons for this: larger than average [Na/Fe] values or higher than average [Fe/H] values leading to a higher $W(NaD)$ value.
As an example, in this panel there are two CN-weak stars that lie significantly above ($\sim$5$\sigma$) the pink shaded region. These stars are 11\_1\_6218 and Scl-0276 and their metallicities as determined here are [Fe/H] = --0.70 and --1.18, respectively.  While these stars are among the most metal-rich in the group, higher metallicity cannot be the entire explanation of the larger $W(NaD)$ as stars Scl-0233 and Scl-0784 have similar $V-V_{HB}$ values, and similar metallicities ([Fe/H] = --1.15 and --1.12, respectively), yet their $W(NaD)$ values place them close to or within the pink shaded regions.  
There are six other stars in the full sample for this metallicity group that show similarly strong $W(NaD)$ values: Scl-0787, 6\_5\_1071, 11\_1\_4528, Scl39, 10\_8\_1236, and Scl92, and their metallicities range from [Fe/H] = --1.59 to --0.67 dex, which encompasses the full range of metallicities in the Scl metal-rich group.  While high dispersion follow-up is needed for confirmation, as an indication of the increased [Na/Fe] required, we show in the top panel of Fig.\ \ref{na}, as a green dashed line, the locus for the solar value of [Na/Fe] at [Fe/H] = --1.25 dex.  Its location suggests that a minority population among the Scl metal-richer stars may have substantially higher, by perhaps 0.5 dex, [Na/Fe] values compared to those for the bulk of the population. 

Turning now to the middle panel of the Fig. \ref{na}, the most remarkable characteristic is again the notably weaker $W(NaD)$ values for the Scl stars compared to those for the GGC NGC~6752.  The spectrum synthesis calculations suggest that this difference requires [Na/Fe] for the Scl stars to be 0.6--0.8 dex lower than the value for the CN-weak stars in the GGC NGC~6752.  As noted above, such a difference in [Na/Fe] between the Scl stars and the CN-weak (first generation) GGC stars is consistent with the results of the high dispersion studies of  \citet{shetrone2003vlt} and \citet{geisler2005sculptor}.  We note also, given the uncertainty in the $W(NaD)$ values, that there is no requirement for the presence of a spread in [Na/Fe] in this sample of Scl stars.  

As discussed above, the location of the Scl stars in the bottom panel of Fig.\ \ref{na} relative to the pink band suggests for these more metal-poor stars a higher overall [Na/Fe] ratio, and potentially larger range in [Na/Fe], compared to the two more metal-rich Scl groups.  We caution against placing too much weight on this outcome given that the uncertainty in the $W(NaD)$ values for these stars is an appreciable fraction of the measured values.  We note, in particular, that the two Scl giants with largest $W(NaD)$ values in this panel are stars Scl-0847 and Scl-0095.  They have virtually identical locations in the panel at $(V-V_{HB}) \approx -2.6$ and $W(NaD) \approx$ 0.85\AA\/  and their metallicities are [Fe/H] = --2.21 and  [Fe/H] = --2.07, respectively. 

In summary, we conclude that the study of Scl $W(NaD)$ line strengths in our large sample of Scl giants, as discussed in this section, reveals three results. First, that there is no evidence to support any connection between CN-strength and enhanced $W(NaD)$ as is seen in the GGC stars.  Second, the majority of Scl stars have weaker $W(NaD)$ strengths than the GGC stars at comparable luminosities, consistent with [Na/Fe] values for the Scl giants that are $\sim$0.6 dex lower than the  approximately solar [Na/Fe] values for the GGC first generation stars (and the halo field in general).  Both of these results are consistent with earlier high dispersion spectroscopic samples of small samples of Scl giants \citep{shetrone2003vlt,geisler2005sculptor}.  Third, we have found indications of the existence of a minority population ($\sim$5\%) of Scl giants that appear to have significantly higher [Na/Fe] values compared to the majority: high dispersion spectroscopy of these stars is required to confirm our results and to ascertain whether any other elements also show differences from the population norm.

%
%

\begin{table}
\caption{Details of synthesis [Na/Fe]. }
\resizebox{\linewidth}{!}
{\begin{threeparttable}
\renewcommand{\TPTminimum}{\linewidth}
\centering 
\begin{tabular}{l c c c c r} 
\hline 
\hline
\normalsize{} & \normalsize{Lower limit } & \normalsize{Upper limit} & \normalsize{[Fe/H]\tnote{a}} & \\ [0.5ex] 
\normalsize{} & \normalsize{ [Na/Fe]} & \normalsize{ [Na/Fe]}  & \\ [0.5ex] 
\hline 

NGC 288					&	--0.1	& 0.4	 	& --1.32	&\\
NGC 6752					&	0.1		& 0.6 	    & --1.54	&\\
M 55						&	--0.5	& 0.0	  	& --1.94	&\\
 & & & & & \\
Scl [Fe/H]$\geq$-1.6 	&	--1.0 	& --0.5  	& --1.25 $\&$ --1.55 &\\
Scl -1.9$<$[Fe/H]$<$-1.6	&	--1.0	& --0.5	 	& --1.63 $\&$ --1.68	 &\\
Scl [Fe/H]$\leq$-1.9		&	--1.0	& --0.5  	& --1.90 $\&$ --1.95 &\\

\hline 
\end{tabular}
\label{tabla_na_ab} 

\begin{tablenotes}
\item [a] Metallicities assumed in generating the synthesis spectra.
\end{tablenotes}

\end{threeparttable}}
\end{table}

\section{Summary and Conclusion}
\label{s_concl}

In this work we have analysed intermediate resolution blue and red spectra of an extensive sample of red giants in the Scl dwarf spheroidal galaxy in order to investigate whether there is evidence for the existence in Scl of the light-element abundance variations that are well established in GGCs.  As part of the analysis we have also employed similar spectra of red giants in a number of GGCs as comparator objects.  We have used the red spectra, which originate from the AAT/AAOmega instrument, to generate individual metallicity estimates for the Scl stars using a line-strength calibration derived from the GGC spectra.  We find that our Scl metallicity estimates are in good accord with existing determinations for the stars in common.  This enabled us to split the Scl sample into three metallicity groups: low, intermediate and high, each of which has a comparator GGC: M55, NGC~6752 and NGC~288, respectively.

We then measured, on blue spectra for the Scl stars that are from Gemini GMOS-S observations, the CN-band strength indices $S(3839)$ and $S(4142)$, as well as the CH-band strength index $W(G)$.  Similarly indices were determined from the AAOmega blue spectra for the GGC comparison stars.  In addition, we measured the strength of the sodium D-lines, denoted by $W(NaD)$, on the AAOmega red spectra of both the Scl and GGC stars.  Analysis of the GGC-star line and band-strength indices generated results in accord with expectations, given the established existence of the light-element abundance variations in the comparison GGCs.  Specifically, as is evident in Fig.\ \ref{wg_cn_clusters} and Fig.\ \ref{hist2}, the distribution of the CN-band strength index $S(3839)$ in each cluster is bimodal, allowing classification of stars as `CN-weak' or `CN-strong'.  

The CN-strong GGC-stars generally have lower values of the CH-band strength index $W(G)$ and, consequently, the differential indices $\delta S3839$ and $\delta W(G)$ are found to be anti-correlated.  The CN-strong stars also have larger values of the sodium line-strength index $W(NaD)$ compared to CN-weak stars of similar luminosity.  These results are consistent with the standard interpretation that the CN-strong stars have higher nitrogen and sodium abundances, and lower carbon abundances, compared to the CN-weak stars.  Spectrum synthesis calculations confirm this interpretation -- the CN-strong and CN-weak index values are consistent with abundance differences of $\Delta$[N/Fe] $\approx$ +1.0 and $\Delta$[C/Fe] $\approx$ --0.5 for an assumed $\Delta$[O/Fe] $\approx$ --0.4 dex (see Fig.\ \ref{wg_cn_clusters} and Table \ref{variations}).  Moreover, spectrum synthesis supports an abundance difference $\Delta$[Na/Fe] $\approx$ 0.5 dex between the two populations, consistent with the range of [Na/Fe] values seen in these clusters based on high dispersion spectroscopy \citep[e.g.,][]{carretta2009anticorrelation}.

The situation for the Scl stars, however, is more complex yet there is little evidence to support the existence of similar correlated abundance differences to those seen in the GGCs.  This is consistent with the results of \citet{geisler2005sculptor}, who showed from high-dispersion spectra that a small sample of Scl giants lacked the O-Na anti-correlation that is one of the characteristic signatures of the GGC light-element abundance variations phenomenon.  \citet{norris2017populations} reached a similar conclusion for the Carina dSph.  Nevertheless, variations in the CN- and CH-band strength indices are present in our Scl sample, although within each metallicity group there is no evidence that this variation is driven by variations in [Fe/H] -- see Fig.\ \ref{comp_met}.  We also find, in contrast to the negative correlations seen for the GGCs, 
that the differential indices  
$\delta S3839$ and $\delta W(G)$ are positively (for the two metal-poorer groups) or uncorrelated (metal-rich group) for the Scl stars. An analogous analysis of the $\delta S3839$ and $\delta CH$ indices for Scl stars given in \citet{lardo2016carbon} generates a very similar result. The overall range in the indices is comparable with, or perhaps larger than, the range seen for the corresponding GGCs.  This suggests that the range in [N/Fe] and [C/Fe] values in the Scl is similar to those in the GGCs.  However, in each metallicity case, the $S(3839)$ and 
$W(G)$ values are below the values for the GGC stars.  Given that the Scl stars all lie near the RGB-tip, this difference may be the result of larger C-depletions from evolutionary mixing.  Overall, such an interpretation is consistent with both the spectrum synthesis analysis of a representative sub-sample of our stars, for which our [C/Fe] and [N/Fe] abundances are in excellent accord with those of \citet{lardo2016carbon}, and their interpretation that [C/Fe] decreases with increasing luminosity on the Scl RGB at all metallicities, and that there is evidence for a significant dispersion in [C/Fe] at given [Fe/H] among the Scl stars.  

As regards sodium in the Scl stars, the strongest result is the indication that the Scl stars have significantly lower [Na/Fe] values, by of order 0.5 dex, compared to the CN-weak GGC stars, i.e., the first generation GGC stars whose abundance ratios are similar to those for field halo stars at similar metallicity.  This result has been seen before in dSph galaxies: \citet{shetrone2003vlt} and \citet{geisler2005sculptor} revealed the [Na/Fe] deficiency in Scl based on high dispersion spectra of a small sample of Scl giants.  Further, \citet{norris2017populations} has shown the same situation applies in Carina, and the extensive studies of \citet{letarte2010} and \citet{lemasle2014} show that this is also the case in the Fornax dSph.
Similarly,  \citet{hasselquist2017apogee} have used near-IR APOGEE spectra to study in detail the abundance patterns in a large sample of stars in the Sgr dSph.  While the sample is dominated by relatively metal-rich stars in the Sgr core, a deficiency in [Na/Fe] of size $\sim$0.4 dex compared to Milky Way stars of similar [Fe/H] values is evident.  \citet{hasselquist2017apogee}  \citep[see also][and the references therein]{mcwilliam2013chemistry} argue that the deficiency in [Na/Fe], along with the overall abundance patterns in their Sgr sample, suggests that the initial mass function for the star formation in Sgr was `top-light', i.e., relatively lacking in massive stars that are the primary nucleosynthetic source for sodium.  This may also be the explanation for the relative deficiency in [Na/Fe] in Scl and the other dSphs.  Intriguingly, we do see evidence for a small population ($\sim$5\%) of Scl giants that appear to have notably higher [Na/Fe] ratios than the bulk of the Scl population.  Such a population is not evident in the Carina sample of \citet{norris2017populations} but inspection of Fig.\ 6 of \citet{hasselquist2017apogee} suggests that there are also a small number of Sgr giants (at lower [Fe/H] than the bulk of the sample) that have relatively high [Na/Fe] values.  Such stars, and the objects in Scl, deserve further attention.

In conclusion, our principle result is that the typical signatures of the GGC light element abundance anomalies are not present in our Sculptor data set, indicating that the dSph star formation environment must be fundamentally different to that which gives rise to the multiple populations in GGCs.  For completeness, we note that the Milky Way halo (as distinct from the Bulge \citep{schiavon2016chemical}) has small population ($\sim$3\%) of stars with CN/CH indices like those for GGCs stars \citep{martell2011building}.  With our sample size we cannot rule a similar frequency of such stars in Scl, but there is no evidence that Scl has possessed a globular cluster whose dissolution would be the origin of such stars, in the same way as is hypothesised for the halo field objects.  \citet{lardo2016carbon} reached a similar conclusion.   
 
There are two fundamental differences between the Scl dSph and the comparison GGCs that may be relevant as to why the light element abundance anomalies are seen in the GGCs and not in Scl and other dSphs.  The first is that the Scl dSph is dark matter dominated \citep[e.g.][]{walker2009universal}, while GGCs are generally considered dark-matter free.  The second difference is structural in the sense that length scales, e.g.\ half-mass radii, are substantially larger in dSphs than for GGCs, and correspondingly, the central densities are much higher in GGCs than in dSphs.  Given these differences, it is probable that GGCs form at the centres of large star-forming complexes and with high SF rates, while dSphs like Scl are likely the result of independent evolution within a dark-matter halo with relatively low star-formation rates.  While the actual mechanism generating the light element abundance anomalies remains unknown, it would seem that high stellar density and high star formation efficiency at formation are essential requirements.  Exploration of the presence or absence of the exact same GGC light element abundance anomalies in the SMC 6-8 Gyr old clusters \citep{hollyhead2016evidence, hollyhead2018kron} is therefore a vital task, as these clusters effectively differ only in age from the GGC\@.  A definitive outcome would decide, for example, whether `special conditions at the earliest epochs' are required, or not. 

\begin{table*}
\caption{Observational data for the 45 stars in the Sculptor dSph galaxy observed with GMOS-S and AAOmega\@. IDs in the form nn\_n\_nnnn are from \citet{coleman2005absence}, Scl-nnnn correspond to stars from \citet{walker2009stellar}, and Sclnn are from \citet{battaglia2008analysis}. Some S(3839) values are missing due to the inter-chip gaps in GMOS-S affecting the spectra.  Missing W(NaD) and W(Ba) values result from low S/N for the fainter stars, while missing [Fe/H] values are a consequence of reduced wavelength coverage for fibres at the edges of the AAOmega camera field in the red spectra. }
\resizebox{\linewidth}{!}
{\begin{threeparttable}
\renewcommand{\TPTminimum}{\linewidth}
\centering 
\begin{adjustbox}{max width=20cm, totalheight={18cm}}
\begin{tabular}{l c c c c c c c c c c c r}
\hline 
\hline

\normalsize{ID} & \normalsize{RA}      & \normalsize{Dec} & \normalsize{V}   & \normalsize{V-I} &  \normalsize{S(3839)} & \normalsize{S(4142) } &\normalsize{W(G) } & \normalsize{W(NaD)} & \normalsize{W(Ba)} & \normalsize{[Fe/H] } &\\ [0.5ex]
\normalsize{ }  & \normalsize{J2000} & \normalsize{J2000} & \normalsize{mag} & \normalsize{mag} &  \normalsize{mag}     & \normalsize{mag}      &\normalsize{\AA}   & \normalsize{\AA}   &\normalsize{\AA} & \normalsize{dex}     &\\

\hline 
		
11\_1\_4953 & 0:59:28.29  &  -33:42:07.3	&   17.013 &	1.630	&	   0.056	&	-0.204	&	7.95  &	0.805	&	0.161		&	-1.44	&\\
11\_1\_5609 & 0:59:16.93  &  -33:40:10.6	&   17.121 &	1.615	&	   -0.272	&	-0.380	&	8.05  &	0.472	&	0.215		&	-2.00	&\\
10\_8\_2607 & 1:00:16.28  &  -33:42:37.2 	&   17.127 &	1.589	&	   0.181	&	-0.072 	&	13.21 &	0.715	&	1.133		&	-1.39	&(a)&\\
11\_1\_6076 & 0:59:08.59  &  -33:41:52.7 	&   17.230 &	1.618	&	   -0.129	&	-0.266	&	7.31  &	0.610	&	0.196		&	-1.93	&\\
Scl38       & 1:00:12.76  &  -33:41:16.0 	&   17.320 &	1.488	&	   -0.018	&	-0.240	&	7.72  &	0.664	&	0.174		&	-2.05	&\\
10\_8\_3291 & 1:00:06.02  &  -33:39:39.6 	&   17.350 &	1.535	&	   0.004	&	-0.242	&	8.54  &	0.773	&	0.243		&	-1.96	&\\
Scl28       & 1:00:15.37  &  -33:39:06.2 	&   17.382 &	1.536	&	   -		&	-0.201	&	7.29  &	0.559	&	0.196		&	-1.84	&\\
Scl46       & 1:00:23.84  &  -33:42:17.4 	&   17.386 &	1.512	&	   0.016	&	-0.225	&	8.94  &	0.587	&	0.234		&	-1.95	&\\
10\_8\_4149 & 0:59:52.99  &  -33:39:18.9 	&   17.454 &	1.560	&	   -0.051	&	-0.283	&	11.66 &	0.627	&	0.192		&	-1.35	&\\
Scl17       & 0:59:56.60  &  -33:36:41.7 	&   17.522 &	1.436	&	   -0.092	&	-0.236	&	8.44  &	0.462	&	0.196		&	-1.95	&\\
11\_1\_6268 & 0:59:04.32  &  -33:44:05.8	&   17.551 &	1.437	&	   -0.220	&	-0.328	&	8.73  &	0.569	&	0.212		&	-1.80	&\\
Scl-0233    & 0:59:14.56  &  -33:40:40.0 	&   17.552 &	1.507	&	   -0.090	&	-0.319	&	9.59  &	0.699	&	0.172		&	-1.15	&\\
10\_8\_2179 & 1:00:22.97  &  -33:43:02.2	&   17.559 &	1.561	&	   -0.064	&	-0.252	&	9.61  &	0.590	&	0.165		&	-1.35	&\\
Scl-0268    & 1:00:07.57  &  -33:37:03.8	&   17.576 &	1.420	&	   -0.230	&	-0.198	&	8.25  &	0.486	&	0.243		&	-1.69	&\\
Scl-0272    & 1:00:04.62  &  -33:41:12.0  	&   17.584 &	1.393	&	   -0.150	&	-0.296	&	6.01  &	0.375	&	0.131		&	-2.50	&\\
Scl-0248    & 0:59:40.32  &  -33:36:06.6  	&   17.585 &	1.438	&	   -0.195	&	-0.234	&	9.17  &	-	&	-		&	-	&\\
11\_1\_6218 & 0:59:05.35  &  -33:42:23.8 	&   17.593 &	1.616	&	   -0.145	&	-0.237	&	11.86 &	0.926	&	0.240		&	-0.70	&\\
Scl-0784    & 0:59:47.25  &  -33:46:30.5  	&   17.598 &	1.510	&	   -0.053	&	-0.148	&	9.82  &	0.736	&	0.297		&	-1.12	&\\
11\_1\_5906 & 0:59:11.83  &  -33:41:25.3	&   17.599 &	1.425	&	   -0.110	&	-0.324	&	7.90  &	0.462	&	0.224		&	-2.02	&\\
Scl-0246    & 0:59:38.11  &  -33:35:08.1	&   17.602 &	1.408	&	   -0.005	&	-0.235	&	8.14  &	-	&	0.201		&	-2.36	&\\
Scl-0468    & 1:00:18.29  &  -33:42:12.2	&   17.610 &	1.435	&	   -0.091	&	-0.250	&	8.15  &	0.479	&	0.206		&	-1.93	&\\
Scl-0247    & 0:59:37.74  &  -33:36:00.1  	&   17.632 &	1.444	&	   0.043	&	-0.214	&	9.63  &	0.504	&	0.212		&	-1.70	&\\
Scl-0276    & 1:00:03.59  &  -33:39:27.1	&   17.681 &	1.311	&	   -0.282	&	-0.344	&	10.70 &	0.918	&	0.244		&	-1.18	&\\
Scl-0492    & 1:00:28.10  &  -33:42:34.4	&   17.684 &	1.366	&	   -0.215	&	-0.315	&	7.54  &	0.558	&	0.189		&	-1.79	&\\
Scl-1310    & 0:59:18.85  &  -33:42:17.5	&   17.690 &	1.460	&	   -0.039	&	-0.300	&	9.57  &	0.680	&	0.195		&	-1.57	&\\
Scl25       & 0:59:41.40  &  -33:38:47.0	&   17.700 &	1.447	&	   -		&	-0.354	&	10.47 &	0.643	&	0.191		&	-1.48	&\\
Scl20       & 0:59:37.22  &  -33:37:10.5 	&   17.715 &	1.500	&	   0.074	&	-0.337	&	10.32 &	0.591	&	0.350		&	-1.27	&\\
Scl-1490    & 1:00:07.44  &  -33:43:19.9 	&   17.720 &	1.377	&	   -0.235	&	-0.250	&	7.23  &	0.431	&	0.148		&	-2.02	&\\
Scl-0216    & 0:59:15.14  &  -33:42:54.7	&   17.732 &	1.389	&	   -0.237	&	-0.236	&	8.33  &	0.466	&	0.248		&	-1.42	&\\
Scl-1020    & 0:59:58.25  &  -33:41:08.6	&   17.750 &	1.420	&	   0.203	&	-0.274	&	10.10 &	0.571	&	0.177		&	-1.71	&\\
Scl-0646    & 1:00:02.55  &  -33:48:49.9  	&   17.755 &	1.382	&	   -		&	-0.303	&	7.10  &	0.590	&	0.237		&	-2.05	&\\
Scl-0263    & 0:59:49.21  &  -33:39:48.9  	&   17.762 &	1.349	&	   -		&	-0.296	&	6.96  &	0.511	&	0.145		&	-2.01	&\\
Scl-0470    & 1:00:17.35  &  -33:41:08.4  	&   17.784 &	1.452	&	   -0.009	&	-0.411	&	10.42 &	0.649	&	0.185		&	-1.32 &\\
Scl-0639    & 0:59:58.24  &  -33:45:50.7	&   17.793 &	1.409	&	   -0.349	&	-0.376	&	9.28  &	-	&	-		&	-	&\\
Scl10       & 0:59:40.46  &  -33:35:53.8 	&   17.826 &	1.360	&	   -0.131	&	-0.310	&	6.67  &	0.451	&	0.194		&	-2.36	&\\
Scl23       & 0:59:54.47  &  -33:37:53.4	&   17.843 &	1.236	&	   -0.285	&	-0.357	&	9.34  &	0.590	&	0.198		&	-1.65	&\\
Scl-1036    & 1:00:39.23  &  -33:42:12.8 	&   17.852 &	1.302	&	   -0.195	&	-0.330	&	5.05  &	0.440	&	0.224		&	-2.24	&\\
Scl24       & 0:59:57.60  &  -33:38:32.5	&   17.872 &	1.189	&	   -0.146	&	-0.331	&	8.68  &	0.536	&	0.197		&	-1.99	&\\
Scl-0437    & 1:00:11.79  &  -33:42:16.9	&   17.886 &	1.449	&	   -0.146	&	-0.369	&	10.12 &	0.534	&	0.236		&	-1.24	&\\
Scl81       & 1:00:06.98  &  -33:47:09.7 	&   17.932 &	1.343	&	  -0.201	&	-0.303	&	7.78  &	0.691	&	0.264		&	-2.07	&\\
Scl13       & 0:59:30.44  &  -33:36:05.0 	&   18.031 &	1.260	&	   -0.126	&	-0.278	&	6.71  &	0.396	&	0.105		&	-	&\\
Scl74       & 1:00:05.93  &  -33:45:56.5 	&   18.041 &	1.291	&	  -0.220	&	-0.245	&	6.27  &	-	&	-		&	-	&\\
Scl59       & 1:00:17.36  &  -33:43:59.6 	&   18.054 &	1.319	&	   -0.053	&	-0.272	&	6.88  &	0.354	&	0.196		&	-2.13	&\\
Scl54       & 1:00:15.18  &  -33:43:11.0 	&   18.087 &	1.397	&	   0.031	&	-0.285	&	10.07 &	0.647	&	0.192		&	-1.41	&\\
Scl49       & 0:59:08.60  &  -33:42:29.4 	&   18.113 &	1.219	&	   -0.214	&	-0.280	&	6.22  &	-	&	-		&	-	&\\

\hline 
\end{tabular}
\end{adjustbox}
\label{tabla_total} 
\begin{tablenotes}
\item (a) This star is the CEMP-s star discussed in \citet{salgado2016scl}.  In that paper the [Fe/H] value of --1.0 from \citet{geisler2005sculptor} was adopted.
\end{tablenotes}

\end{threeparttable}}
\end{table*}


\begin{table*}
\caption{Observational data for the 108 stars in the Sculptor dSph galaxy observed only with AAOmega\@. IDs use the same nomenclature as in Table \ref{tabla_total}.  Missing W(NaD) and W(Ba) values result from low S/N for the fainter stars, while missing [Fe/H] values are a consequence of reduced wavelength coverage for fibres at the edges of the AAOmega camera field in the red spectra. }
\resizebox{\linewidth}{!}
{\begin{threeparttable}
\renewcommand{\TPTminimum}{\linewidth}
\centering 
\begin{adjustbox}{max width=20cm,totalheight={22.5cm}}
\begin{tabular}{l c c c c c c c c c r}
\hline 
\hline	 

\normalsize{ID} & \normalsize{RA}      & \normalsize{Dec}     &  \normalsize{V}  & \normalsize{V-I} & \normalsize{W(NaD) }  & \normalsize{W(Ba)} & \normalsize{[Fe/H] }  &\\ [0.5ex]
\normalsize{ }  & \normalsize{J2000} & \normalsize{J2000} &  \normalsize{mag}
&  \normalsize{mag} & \normalsize{\AA}      &\normalsize{\AA}    & \normalsize{dex}      &\\

\hline 

6\_6\_0152 		&	  01:01:14.53	&	-33:30:09.2		&		16.862	&	1.587		&		0.782	&	0.298		&	-1.712		\\
Scl04			&	  00:59:55.63	&	-33:33:24.6		&		16.890	&	1.297		&		0.613	&	0.197		&	-2.406		\\
10\_8\_0644		&	  01:01:02.87	&	-33:38:52.2		&		17.033	&	1.562		&		0.686	&	0.194		&	-2.077		\\
11\_1\_6226 	&	  00:59:04.80	&	-33:38:44.2		&		17.046	&	1.704		&		0.880	&	0.223		&	-1.617		\\
6\_3\_0197		&	  01:02:22.18	&	-33:23:04.4		&		17.108	&	1.482		&		0.690	&	-			&	-2.355		\\
Scl92			&	  01:00:25.30	&	-33:50:50.8		&		17.109	&	1.159		&		0.962	&	0.202		&	-0.671		\\
10\_7\_0022		&	  01:01:49.38	&	-33:54:09.9		&		17.130	&	1.507		&		0.444	&	0.145		&	-2.602		\\
11\_2\_0504 	&	  00:59:43.15	&	-33:56:46.7		&		17.130	&	1.550		&		0.533	&	0.194		&	-2.254		\\
10\_8\_1013		&	  01:00:49.36	&	-33:42:00.5		&		17.137	&	1.587		&		0.663	&	0.197		&	-2.116		\\
6\_5\_0546		&	  01:00:42.55	&	-33:35:47.3		&		17.164	&	1.496		&		0.624	&	0.182		&	-1.780		\\
10\_1\_1614		&	  01:02:23.58	&	-33:43:43.8		&		17.184	&	1.547		&		0.460	&	0.205		&	-2.323		\\
11\_1\_7127		&	  00:58:42.58	&	-33:48:33.4		&		17.189	&	1.576		&		0.721	&	0.215		&	-1.976		\\
Scl27			&	  01:00:34.04	&	-33:39:04.6		&		17.200	&	1.701		&		0.727	&	0.267		&	-1.346		\\
11\_7\_0120		&	  00:57:55.06	&	-33:59:40.9		&		17.201	&	1.449		&		0.709	&	0.224		&	-2.304		\\
Scl84			&	  01:00:38.12	&	-33:48:16.9		&		17.212	&	1.482		&		0.582	&	0.177		&	-2.054		\\
Scl70			&	  01:00:50.87	&	-33:45:05.2		&		17.213	&	1.673		&		0.678	&	0.204		&	-1.568		\\
10\_8\_0128		&	  01:01:39.62	&	-33:45:04.3		&		17.222	&	1.604		&		0.541	&	0.234		&	-1.807		\\
7\_2\_0287		&	  00:58:45.46	&	-33:10:00.4		&		17.223	&	1.477		&		0.640	&	0.304		&	-1.383		\\
7\_6\_0038		&	  00:58:15.72	&	-33:27:16.2		&		17.226	&	1.454		&		0.711	&	0.188		&	-1.943		\\
11\_1\_4989 	&	  00:59:27.68	&	-33:40:35.6		&		17.229	&	1.475		&		0.556	&	0.166		&	-2.163		\\
Scl77			&	  00:59:55.68	&	-33:46:40.1		&		17.233	&	1.388		&		0.690	&	0.191		&	-2.179		\\
6\_5\_0027		&	  01:01:49.33	&	-33:36:23.9		&		17.233	&	1.505		&		0.582	&	0.216		&	-1.623		\\
11\_8\_0435		&	  00:56:46.94	&	-33:48:20.6		&		17.235	&	1.480		&		-		&	-			&	-2.048		\\
10\_8\_0856		&	  01:00:54.17	&	-33:40:14.6		&		17.257	&	1.538		&		0.647	&	0.282		&	-1.843		\\
10\_3\_0033		&	  01:03:53.61	&	-34:13:47.9		&		17.261	&	1.614		&		0.691	&	0.165		&	-2.177		\\
Scl76 			&	  00:59:12.09	&	-33:46:20.8		&		17.266	&	1.386		&		0.561	&	0.152		&	-2.004		\\
11\_1\_5108		&	  00:59:26.25	&	-33:46:52.9		&		17.272	&	1.541		&		0.679	&	0.111		&	-1.899		\\
10\_8\_1236		&	  01:00:42.48	&	-33:44:23.4		&		17.273	&	1.581		&		0.956	&	0.217		&	-1.595		\\
Scl39			&	  00:59:46.41	&	-33:41:23.5		&		17.304	&	1.605		&		0.902	&	0.221		&	-0.956		\\
11\_1\_5411		&	  00:59:20.80	&	-33:44:04.9		&		17.309	&	1.470		&		0.555	&	0.204		&	-2.269		\\
11\_1\_4528		&	  00:59:35.36	&	-33:44:09.4		&		17.369	&	1.672		&		0.859	&	0.288		&	-1.220		\\
11\_5\_0233		&	  00:57:33.46	&	-34:24:47.7		&		17.373	&	1.364		&		0.707	&	0.200		&	-2.115		\\
10\_7\_0734		&	  01:00:18.31	&	-33:53:31.5		&		17.399	&	1.487		&		0.595	&	-			&	-1.897		\\
6\_5\_1071 		&	  01:00:17.76	&	-33:35:59.6		&		17.400	&	1.549		&		0.819	&	0.247		&	-1.289		\\
6\_5\_0436		&	  01:00:50.23	&	-33:36:38.2		&		17.412	&	1.403		&		0.609	&	0.151		&	-2.147		\\
6\_6\_0203		&	  01:00:54.64	&	-33:25:23.0		&		17.424	&	1.503		&		0.620	&	0.160		&	-1.826		\\
7\_5\_0227		&	  00:57:32.70	&	-33:34:06.9		&		17.428	&	1.362		&		0.586	&	0.141		&	-1.992		\\
6\_5\_0153		&	  01:01:21.65	&	-33:36:19.5		&		17.470	&	1.326		&		0.502	&	0.117		&	-2.667		\\
6\_6\_0338 		&	  01:00:18.20	&	-33:31:40.4		&		17.473	&	1.597		&		0.752	&	0.223		&	-1.467		\\
10\_7\_0487		&	  01:00:45.63	&	-34:01:27.5		&		17.480	&	1.430		&		0.538	&	0.322		&	-2.202		\\
10\_8\_2693		&	  01:00:14.82	&	-33:44:22.1		&		17.481	&	1.468		&		0.560	&	0.148		&	-2.009		\\
Scl09			&	  01:00:20.29	&	-33:35:34.5		&		17.495	&	1.409		&		0.561	&	-			&	-2.085		\\
10\_7\_0815		&	  01:00:10.92	&	-34:01:34.8		&		17.496	&	1.386		&		0.488	&	0.142		&	-2.179		\\
Scl88			&	  01:00:00.49	&	-33:49:35.8		&		17.512	&	1.365		&		0.466	&	0.142		&	-2.556		\\
10\_7\_0891		&	  01:00:04.20	&	-33:52:21.8		&		17.525	&	1.364		&		0.437	&	0.096		&	-2.544		\\
11\_1\_7421		&	  00:58:33.76	&	-33:43:18.6		&		17.525	&	1.449		&		0.665	&	0.264		&	-1.888		\\
11\_1\_5592		&	  00:59:17.78	&	-33:46:01.7		&		17.526	&	1.466		&		0.501	&	0.184		&	-1.715		\\
11\_1\_5483		&	  00:59:19.15	&	-33:38:51.0		&		17.546	&	1.415		&		0.566	&	-			&	-1.779		\\
Scl-0259 		&	  00:59:54.20	&	-33:40:27.0		&		17.560	&	1.520		&		0.743	&	-			&	-1.506		\\
Scl-0665		&	  01:00:44.27	&	-33:49:18.8		&		17.570	&	1.480		&		0.561	&	0.177		&	-1.565		\\
Scl-0993		&	  01:00:08.98	&	-33:50:22.9		&		17.590	&	1.370		&		0.664	&	0.209		&	-2.034		\\
11\_2\_0913		&	  00:58:46.71	&	-33:53:36.8		&		17.592	&	1.375		&		0.598	&	0.270		&	-1.828		\\
Scl-0782 		&	  00:59:39.48	&	-33:45:39.2		&		17.610	&	1.410		&		0.432	&	-			&	-1.537		\\
10\_1\_0634		&	  01:03:18.66	&	-33:47:15.8		&		17.622	&	1.506		&		0.560	&	0.203		&	-1.846		\\

\hline 
\end{tabular}
\end{adjustbox}
\label{tabla_total2} 

\end{threeparttable}}
\end{table*}


\begin{table*}
\contcaption{A table continued from the previous one.}
\label{tab:continued}
\resizebox{\linewidth}{!}
{\begin{threeparttable}
\renewcommand{\TPTminimum}{\linewidth}
\centering 
\begin{adjustbox}{max width=20cm, totalheight={22.5cm}}
\begin{tabular}{l c c c c c c c c r}
\hline 
\hline		 

\normalsize{ID} & \normalsize{RA}      & \normalsize{Dec}     &  \normalsize{V} & \normalsize{V-I} & \normalsize{W(NaD) }  & \normalsize{W(Ba)} & \normalsize{[Fe/H] }  &\\ [0.5ex]
\normalsize{ }  & \normalsize{J2000} & \normalsize{J2000} &  \normalsize{mag} &  \normalsize{mag} & \normalsize{\AA}      &\normalsize{\AA}    & \normalsize{dex}      &\\

\hline 

Scl-0108		&	  01:01:19.27	&	-33:45:41.6		&		17.630	&	1.420		&		0.532	&	0.221		&	-1.992		\\
Scl-0968		&	  00:59:14.37	&	-33:31:43.8		&		17.640	&	1.340		&		0.536	&	0.223		&	-1.583		\\
Scl-1183		&	  00:59:42.57	&	-33:42:18.2		&		17.650	&	1.500		&		0.543	&	0.202		&	-1.257		\\
Scl-0822		&	  00:59:44.11	&	-33:28:11.7		&		17.680	&	1.360		&		0.593	&	0.273		&	-1.449		\\
7\_4\_1809		&	  00:59:47.21	&	-33:33:37.0		&		17.680	&	1.336		&		0.508	&	0.186		&	-2.198		\\
Scl-0674		&	  01:00:48.22	&	-33:49:55.5		&		17.720	&	1.400		&		0.659	&	-			&	-2.298		\\
Scl-0636		&	  01:00:53.68	&	-33:52:27.4		&		17.720	&	1.380		&		0.490	&	0.109		&	-1.904		\\
Scl-0772		&	  00:59:46.15	&	-33:48:39.1		&		17.720	&	1.440		&		0.813	&	0.264		&	-		\\
10\_2\_0586		&	  01:02:08.04	&	-33:57:50.4		&		17.732	&	1.392		&		0.542	&	0.108		&	-1.833		\\
11\_2\_0748		&	  00:59:13.96	&	-34:00:38.5		&		17.737	&	1.292		&		0.570	&	0.313		&	-2.168		\\
Scl-0667		&	  01:00:29.22	&	-33:55:44.2		&		17.740	&	1.400		&		0.619	&	0.149		&	-1.709		\\
Scl-1281		&	  00:59:36.98	&	-33:30:28.3		&		17.740	&	1.310		&		0.621	&	-			&	-1.679		\\
Scl-0649		&	  01:00:18.96	&	-33:45:14.7		&		17.750	&	1.300		&		0.571	&	0.200		&	-2.084		\\
Scl-0242		&	  00:59:37.77	&	-33:41:15.4		&		17.750	&	1.370		&		0.636	&	0.160		&	-1.745		\\
Scl-0454		&	  01:00:25.78	&	-33:30:25.4		&		17.760	&	1.320		&		0.463	&	0.212		&	-2.133		\\
Scl-0655		&	  01:00:21.11	&	-33:56:27.9		&		17.770	&	1.320		&		0.425	&	-			&	-2.258		\\
Scl-0892		&	  01:01:45.19	&	-33:25:40.1		&		17.770	&	1.300		&		0.673	&	0.205		&	-2.287		\\
Scl-0847		&	  01:00:38.91	&	-33:26:09.6		&		17.780	&	1.350		&		0.842	&	-			&	-2.216		\\
Scl-0095		&	  01:01:15.09	&	-33:42:41.8		&		17.780	&	1.350		&		0.867	&	-			&	-2.074		\\
Scl-0790		&	  00:59:48.16	&	-33:50:22.7		&		17.790	&	1.420		&		0.717	&	0.238		&	-0.722		\\
Scl-0947		&	  00:58:38.77	&	-33:35:02.3		&		17.800	&	1.300		&		0.609	&	0.192		&	-2.253		\\
Scl50			&	  00:59:47.05	&	-33:42:54.2		&		17.810	&	1.389		&		0.586	&	0.281		&	-1.541		\\
Scl-0472		&	  01:00:27.01	&	-33:38:22.0		&		17.820	&	1.420		&		0.621	&	-			&	-1.609		\\
Scl89			&	  01:00:10.49	&	-33:49:36.9		&		17.829	&	1.296		&		0.524	&	0.137		&	-1.909		\\
Scl-0099		&	  01:00:48.54	&	-33:40:52.7		&		17.830	&	1.420		&		0.770	&	-			&	-1.484		\\
10\_1\_2031		&	  01:01:59.91	&	-33:51:12.9		&		17.861	&	1.407		&		0.639	&	-			&	-1.865		\\
Scl-0653 		&	  01:00:24.62	&	-33:44:28.9		&		17.870	&	1.340		&		0.617	&	-			&	-1.772		\\
Scl-0804		&	  01:01:30.79	&	-33:57:29.9		&		17.870	&	1.340		&		0.691	&	-			&	-2.524		\\
Scl-0456		&	  01:00:26.19	&	-33:31:38.8		&		17.870	&	1.300		&		0.529	&	0.146		&	-2.180		\\
Scl-0785 		&	  00:59:47.66	&	-33:47:29.4		&		17.880	&	1.460		&		0.561	&	0.194		&	-1.315		\\
Scl-0874		&	  00:58:04.07	&	-33:50:41.8		&		17.880	&	1.380		&		0.615	&	0.111		&	-2.025		\\
Scl-0482		&	  01:00:37.78	&	-33:44:08.7		&		17.880	&	1.340		&		0.566	&	0.228		&	-2.008		\\
Scl-0657		&	  01:00:33.85	&	-33:44:54.4		&		17.890	&	1.340		&		0.445	&	-			&	-1.822		\\
Scl-0289 		&	  00:58:46.95	&	-33:38:53.5		&		17.890	&	1.350		&		0.608	&	0.201		&	-2.098		\\
Scl-0522		&	  01:02:25.28	&	-33:39:50.0		&		17.890	&	1.310		&		0.572	&	0.216		&	-2.228		\\
Scl-1140		&	  01:00:11.73	&	-33:44:50.3		&		17.900	&	1.320		&		0.313	&	0.126		&	-2.461		\\
Scl-0683 		&	  01:00:01.44	&	-33:51:16.7		&		17.900	&	1.290		&		-		&	-			&	-1.731		\\
Scl80			&	  01:00:42.96	&	-33:47:06.4		&		17.907	&	1.138		&		0.533	&	0.178		&	-2.552		\\
Scl-1034		&	  01:00:45.42	&	-33:52:14.7		&		17.910	&	1.260		&		0.717	&	0.441		&	-1.523		\\
Scl-1321		&	  00:59:49.90	&	-33:44:05.0		&		17.910	&	1.340		&		0.616	&	0.164		&	-1.610		\\
Scl-0509		&	  01:02:18.74	&	-33:36:15.3		&		17.910	&	1.250		&		-		&	-			&	-2.474		\\
Scl-1054		&	  00:57:35.52	&	-33:41:21.9		&		17.940	&	1.320		&		0.754	&	0.198		&	-1.780		\\
Scl-0264		&	  00:59:59.09	&	-33:36:44.9		&		17.950	&	1.280		&		0.539	&	0.142		&	-2.489		\\
Scl-0106		&	  01:01:10.27	&	-33:38:37.8		&		17.950	&	1.250		&		0.607	&	0.101		&	-2.109		\\
Scl-0669		&	  01:00:46.51	&	-33:46:47.1		&		17.960	&	1.300		&		0.568	&	0.171		&	-2.568		\\
Scl-0668		&	  01:00:34.32	&	-33:49:52.8		&		17.960	&	1.420		&		0.693	&	0.159		&	-1.286		\\
Scl-0787		&	  00:59:44.04	&	-33:49:41.1		&		17.960	&	1.370		&		0.960	&	0.272		&	-1.420		\\
Scl-1287		&	  00:59:37.61	&	-33:40:22.0		&		17.960	&	1.330		&		0.521	&	-			&	-2.037		\\
Scl-0223		&	  00:59:11.34	&	-33:37:28.2		&		17.980	&	1.330		&		0.607	&	0.130		&	-1.668		\\
Scl-0476		&	  01:00:32.83	&	-33:36:07.4		&		17.980	&	1.280		&		0.422	&	0.169		&	-1.919		\\
Scl26 			&	  00:59:30.49	&	-33:39:04.0		&		17.981	&	1.252		&		0.527	&	0.200		&	-2.094		\\
Scl07			&	  00:59:58.31	&	-33:34:40.4		&		17.983	&	1.283		&		0.583	&	0.233		&	-1.347		\\
Scl-0890		&	  01:01:00.91	&	-33:27:54.9		&		17.990	&	1.310		&		0.492	&	-			&	-1.828		\\
Scl-0835		&	  01:00:02.68	&	-33:30:25.1		&		18.000	&	1.310		&		0.502	&	0.162		&	-			\\

\hline 
\end{tabular}
\end{adjustbox}
\label{tabla_total3} 

\end{threeparttable}}
\end{table*}

\section*{Acknowledgements}
\addcontentsline{toc}{section}{Acknowledgements}

Based in part on observations (Program GS-2012B-Q-5) obtained at the Gemini Observatory, which is operated by the Association of Universities for Research in Astronomy, Inc., under a cooperative agreement with the NSF on behalf of the Gemini partnership: the National Science Foundation (United States), the National Research Council (Canada), CONICYT (Chile), Ministerio de Ciencia, Tecnolog\'{i}a e Innovaci\'{o}n Productiva (Argentina), and Minist\'{e}rio da Ci\^{e}ncia, Tecnologia e Inova\c{c}\~{a}o (Brazil).
CS acknowledges support provided by CONICYT, Chile, through its scholarships program  CONICYT-BCH/Doctorado Extranjero 2013-72140033. This research has been supported in part by the Australian Research Council through Discovery Projects grants DP120101237, DP150103294 and FT140100554.

\bibliographystyle{mnras}
\bibliography{scl_mnras}

\bsp	
\label{lastpage}
\end{document}